\documentclass{aa}
\bibpunct{(}{)}{;}{a}{}{,}
\usepackage[varg]{txfonts}
\usepackage{graphicx}
\usepackage{subcaption}
\usepackage[normalem]{ulem}
\usepackage{amsmath, amssymb, amsfonts}

\title{Mergers Do Trigger AGNs out to $z\sim0.6$}

\author{F.~Gao\inst{\ref{inst1}}
\and L.~Wang\inst{\ref{inst1},\ref{inst2}}
\and W.~J.~Pearson\inst{\ref{inst1},\ref{inst2}}
\and Y.~A.~Gordon\inst{\ref{inst4}}
\and B.~W.~Holwerda\inst{\ref{inst5}}
\and A.~M.~Hopkins\inst{\ref{inst6}}
\and M.~J.~I.~Brown\inst{\ref{inst7}}
\and J.~Bland-Hawthorn\inst{\ref{inst8}}
\and M.~S.~Owers\inst{\ref{inst9},\ref{inst10}}
}
\institute{Kapteyn Astronomical Institute, University of Groningen, Postbus 800, 9700 AV Groningen, The Netherlands\label{inst1}
\and 
SRON Netherlands Institute for Space Research, Landleven 12, 9747 AD, Groningen, The Netherlands\label{inst2}
\and Department of Physics and Astronomy, University of Manitoba, Winnipeg, MB R3T 2N2, Canada\label{inst4}
\and Department of Physics and Astronomy, 102 Natural Science Building, University of Louisville, Louisville KY 40292, USA\label{inst5}
\and Australian Astronomical Optics, Macquarie University, 105 Delhi Rd, North Ryde, NSW 2113, Australia
\label{inst6}
\and School of Physics and Astronomy, Monash University, Clayton, Victoria 3800, Australia\label{inst7}
\and{Sydney Institute for Astronomy, School of Physics A28, University of Sydney, NSW 2006, Australia}\label{inst8}
\and{Department of Physics and Astronomy, Macquarie University, Sydney, NSW2109, Australia}\label{inst9}
\and{Astronomy, Astrophysics and Astrophotonics Research Centre, Macquarie University, Sydney, NSW 2109, Australia}\label{inst10}
}
\abstract{}{The fueling and feedback of Active Galactic Nuclei (AGNs) are important in understanding the co-evolution between black holes and host galaxies. Mergers are thought to have the capability to bring gas inwards and ignite nuclear activity, especially for more powerful AGNs. However, there is still significant ongoing debate on whether mergers can trigger AGNs and, if they do, whether mergers are a significant triggering mechanism.}{We select a low-redshift ($0.005<z<0.1$) sample from the Sloan Digital Sky Survey (SDSS) and a high-redshift ($0<z<0.6$) sample from the Galaxy And Mass Assembly (GAMA) survey. We take advantage of the convolutional neural network technique to identify mergers. We use mid-infrared (MIR) color cut and optical emission line diagnostics to classify AGNs. We also include Low Excitation Radio Galaxies (LERGs) to investigate the connection between mergers and low accretion rate AGNs.}{We find that AGNs are more likely to be found in mergers than non-mergers, with an AGN excess up to $1.81\pm{0.16}$, suggesting that mergers can trigger AGNs. We also find the fraction of mergers in AGNs is higher than that in non-AGN controls, for both MIR and optically selected AGNs, as well as LERGs, with values between $16.40\pm{0.5}\%$ and $39.23\pm{2.10}\%$, implying a non-negligible to potentially significant role of mergers in triggering AGNs. 
This merger fraction in AGNs increases as stellar mass increases which supports the idea that mergers are more important for triggering AGNs in more massive galaxies. In terms of merger fraction as a function of AGN power we find a positive trend for MIR selected AGNs and a complex trend for optically selected AGNs, which we interpret under an evolutionary scenario proposed by previous studies. In addition, obscured MIR selected AGNs are more likely to be hosted in mergers than unobscured MIR selected AGNs.}{}

\begin{document}
\maketitle
\thispagestyle{empty}
\section{Introduction}

Almost every massive galaxy in the Universe hosts a super massive black hole \citep[SMBH;][]{Richstone}, although most of them are dormant like the one in our Galaxy with an accretion rate $\leq 10^{-8} M_{\odot} \rm yr^{-1}$ \citep{Baganoff}. Despite its small scale compared to the host galaxy, it has long been confirmed that tight correlations exist between the mass of the SMBH and host galaxy properties. For example, black hole (BH) mass correlates with the velocity dispersion of the galaxy bulge \citep[$\rm M_{BH}$-$\sigma_{bulge}$ relation, e.g.,][]{FerrareseandMerritt, Gebhardt}. Black hole mass also correlates with the luminosity (and stellar mass) of the galaxy bulge \citep[e.g.,][]{McLure, McLure02}. These tight relations support a popular scenario in which SMBHs co-evolve with their host galaxies \citep[see][for a review]{KH13}.  Active Galactic Nuclei (AGNs) are rapidly accreting black holes and are proposed to be the link connecting the central engine and the host galaxy. Both AGN activity and cosmic star formation activity reach their peaks at z $\sim$2 \citep{Richards, MD14}, which further supports the co-evolution picture. These vigorous monsters can shape their hosts either through radiation pressure \citep[radiative/quasar mode:][]{Silk, Haehnelt, Vogelsberger} or via radio jets \citep[maintenance/radio mode:][]{Blanford, Croton, Bower, Lister}. 

An important question in AGN research is what processes bring gas inwards and make it lose most of its angular momentum to accrete in the disk, from host galaxy scale ($\sim 10$ kpc) down to the SMBH scale ($\sim 10$ pc), to fuel nuclear activity. 
Early studies of luminous infrared galaxies (LIRGs) and ultra luminous infrared galaxies (ULIRGs) reveal a high fraction of interactions \citep[e.g.,][]{Murphy96, Veilleux02}. Most of these luminous IR galaxies are thought be AGNs \citep[e.g.,][]{Sanders96}, thus leading to a popular explanation in AGN triggering: mergers. In this scenario, the strong gravitational torque funnels gas inwards and triggers the accretion activity around the central black hole as well as accelerates star formation in the bulge, thus connecting the growth of SMBHs and their host galaxies \citep[e.g.,][]{Hopkins06}. Simulations show that mergers are able to transport gas reservoirs to the inner region of galaxies and trigger nuclear activity, reproducing the observed AGN luminosity function \citep{Hopkins08}.

Many efforts have been made to find an observational connection between galaxy mergers and AGNs. However, the results are still mixed. On the one hand, studies focusing on AGN fraction in mergers compared to non-mergers show that mergers are more likely to host AGNs than non-mergers \citep{Ellison11, Lackner, Satyapal, Weston, Donley18, Goulding}. AGN fraction also increases as the separation between galaxy pairs decreases \citep[e.g.,][]{Ellison11, Silverman11, Koss12, Ellison13, Satyapal, Khabiboulline}. On the other hand, studies focusing on merger fraction in AGNs compared to non-AGN controls show conflicting results. Some found that  AGNs reside more frequently in galaxy mergers compared to non-active counterparts, especially for those with higher luminosity, or dust reddened AGNs or even Compton-thick AGNs \citep{Treister, Santini12, Kocevski15, Comerford, Fan, Ellison19}. While others did not find a difference of merger fraction in active and non-active galaxies \citep{Grogin05, Gabor, Cisternas, Kocevski12, Mechtley, Villforth17}, or a dependence on AGN luminosity \citep[e.g.,][]{Hewlett}, suggesting that major mergers are not the dominant mechanism in triggering AGN activity, even for higher luminosity ones. 

Various reasons are thought to be responsible for these conflicting results. Selection bias is perhaps one of the most important factors. In terms of AGN selection, different studies use different selection criteria for AGNs such as mid-infrared (MIR) color selection \citep[e.g.,][]{Satyapal, Donley18, Goulding, Ellison19}, X-ray selection \citep[e.g.,][]{Hasinger08, Kocevski12, Lackner, Hewlett, Secrest}, optical emission line ratios \citep[the so-called BPT diagram,][]{BPT} and radio \citep{Ellison15, Chiaberge, Gordon19}. AGNs selected in different ways may represent different stages in the merger evolutionary scenario \citep[e.g.,][]{Sanders88}. In terms of merger selection, some studies take advantage of spectroscopic redshift surveys to pick up galaxy pairs within a certain distance \citep[e.g.,][]{Ellison11, Satyapal} which are more likely to be early stage mergers, while others focus on images to select morphologically disturbed galaxies \citep[e.g.,][]{Kocevski12, Donley18, Ellison19}. Morphologically selected mergers can select galaxies with signs of disturbance caused by merging but without a visible merging companion. In addition, some studies use small samples which may not cover a wide range of redshifts, stellar masses, luminosities and do not establish a matched control sample for comparison. When identifying mergers, many studies use visual inspection \citep[e.g.,][]{Cisternas, Kocevski15, Ellison19}, which is based on subjective ranking, or fitting S\'ercic profiles \citep[e.g.,][]{Fan, Mechtley}, which relies on careful point source subtraction. Also, simulations show that non-parametric measurements such as Gini and M20 coefficients \citep[e.g.,][]{Villforth17} work well at the first pass and final coalescence stages but fail at other stages \citep{Lotz08}, and are sensitive to mass ratios and gas fractions \citep{Lotz10a, Lotz10b}. Moreover, the time scale of AGN activity may also play an important role. Typically AGN lifetimes are $\sim 10^7-10^8$ years \citep[e.g.,][]{Marconi04} which is quite short compared to that of mergers which can last up to a few Gyrs \citep{Lotz08, Moreno}. This difference in timescale can lead to fewer AGNs being detected in some stages of the merger process. In addition, the time delay between merger events and the triggering of AGN activity as inflowing gas eventually falling into the vicinity of BH would bias towards fewer AGNs being observed \citep[e.g.,][]{Villforth14, Shabala}.

In this work, we select our samples from two spectroscopic surveys, the Sloan Digital Sky Survey \citep[SDSS;][]{York} at lower redshifts and the Galaxy And Mass Assembly \citep[GAMA][]{Driver09} survey at higher redshifts. We take advantage of the deep learning convolutional neural networks (CNN) to identify mergers, using SDSS images for the SDSS sample and Kilo Degree Survey \citep[KiDS;][]{Kids2, Kids1} images for the GAMA sample. CNNs allow us to rapidly classify very large numbers of objects in a consistent and reproducible manner. With upcoming large area surveys, such as \textit{Euclid} \citep{2011arXiv1110.3193L} and the Large Synoptic Survey Telescope \citep{2009arXiv0912.0201L}, which are expected to produce images of billions of images of galaxies, CNNs offer an efficient way to analyze this data. We use a MIR color cut criterion and optical emission lines diagnostics to identify AGNs. We also use a low excitation radio galaxies (LERGs) catalog from \citet{BH12} to study the connection between low accretion rate AGNs and mergers. Our goal is to combine sophisticated merger selection, large samples and multiple AGN selection methods to explore the merger-AGN connection in the low-redshift Universe.

This paper is structured as follows. In Section 2, we describe our sample construction, merger identification method and AGN selection methods. In Section 3, we show our results on the AGN fraction in mergers and non-merger controls, and merger fraction in AGNs and non-AGN controls. We also investigate the dependence of merger fraction on stellar mass and AGN power. Discussions of our results and comparisons with previous work are presented in Section 4. In Section 5 we summarize our work. Throughout this paper, we assume a flat $\Lambda$CDM universe with $\Omega_{\text{M}} = 0.3$, $\Omega_{\Lambda} = 0.7$, and $H_{0}=70 \,\rm km s^{-1} Mpc^{-1}$.

\section{Data and methods}

We select our samples from two major galaxy surveys. For the first sample we use the SDSS DR7 spectroscopic catalog \citep{sdssdr7} covering $\sim 14000$ deg$^2$ at $0.005<z<0.1$ with a magnitude limit of $r<17.77$. The redshift range is limited by the redshift range of the training sample. The second sample comes from the GAMA spectroscopic survey \citep{Driver, Liske} in the three GAMA equatorial fields (G09, G12, G15, totaling 180 deg$^2$) at $0<z<0.6$. The GAMA spectroscopic survey contains $\sim$ 300,000 galaxies with a magnitude limit of $r<19.8$, covering a sky area of $\sim$ 286 deg$^2$. The KiDS survey is an optical imaging survey that covers $\sim$ 1500 deg$^2$, reaching a magnitude limit of $r<25.2$ and a point spread function (PSF) full width at half maximum (FWHM) of $<0.7\arcsec$, compared to $\sim 1.4\arcsec$ median seeing of SDSS images. SDSS provides us with a large sample of galaxies in the local universe. The GAMA survey combined with KiDS (which offers much better imaging quality in terms of depth and angular resolution) allows us to push our study out to higher redshift.

\subsection{Merger Identification using CNN}\label{merger}

The classification of mergers is performed through the deep learning CNN developed in \citet{2019arXiv190810115P, 2019A&A...626A..49P}, based on the SDSS $gri$ images for the SDSS sample and the KiDS $r-$band images for the GAMA sample. We summarize the merger identification method briefly as follows. 

We use the CNN from \citet{2019arXiv190810115P} for SDSS images and \citet{2019A&A...626A..49P} for KiDS images. The SDSS network was trained using 3003 merging galaxies visually identified within Galaxy Zoo 1 \citep{2008MNRAS.389.1179L}, and then visually confirmed by \citet{2010MNRAS.401.1552D, 2010MNRAS.401.1043D}. A further 3003 non-merging galaxies that have the same redshift range ($0.005 < z < 0.1$) and $r$-band magnitude limit ($<$ 17.77) were also selected. 
Unlike the SDSS images, the training sample of the KiDS images is based on a combination of visual classification and morphological disturbance measurement. The KiDS network is divided into four redshift bins: $0.0 < z < 0.15$, $0.15 < z < 0.3$, $0.3 < z < 0.45$ and $0.45 < z < 0.6$. The KiDS-z00 network (in the lowest redshift bin $0.0 < z < 0.15$) was trained using galaxies from the latest KiDS data release 4 \citep{2019A&A...625A...2K}. The merging galaxies were selected to have the weighted fraction of votes identifying the galaxy as a merger above 0.5 from GAMA-KiDS-Galaxy Zoo \citep{2019AJ....158..103H} and are also identified as a merging galaxy using the smoothness and asymmetry non-parametric statistics \citep{Conselice}, totaling 1917 galaxies. A further 1917 non-merging galaxies were selected that did not meet either of these criteria. 

For the higher redshift KiDS networks there are no pre-classified galaxies available for training. Thus, we use the galaxies used to train the KiDS-z00 network and make them fainter and smaller to appear like higher redshift galaxies, randomly selecting one redshift between $0.15 < z \leq 0.30$, one redshift between $0.30 < z \leq 0.45$ and one redshift between $0.45 < z \leq 0.60$ for each galaxy. The apparent magnitude of the galaxy is corrected for the luminosity distances at the assigned redshifts, removing any galaxies that fall below the limiting magnitude of the KiDS survey. For the remaining galaxies, a rotation by a random angle between 0$^{\circ}$ and 360$^{\circ}$, or a skew by a random angle between $\pm 10^{\circ}$ and $\pm 30^{\circ}$ is applied. For each galaxy we randomly select one of the two transformations or no transformation. The images are then re-binned to match the physical resolution of the KiDS survey for the assigned redshifts and Gaussian noise is added, with a standard deviation of the noise in the original image. The images are not corrected for the change in wavelength of the rest-frame emission. The number of merging and non-merging galaxies are then balanced within each redshift bin by randomly removing galaxies of the classification with more objects. This results in 1902, 1870 and 1789 objects in the $0.15 \leq z < 0.30$, $0.30 \leq z < 0.45$ and $0.45 \leq z < 0.60$ redshift bins respectively.


The CNNs used in this work are trained with visually selected galaxy mergers and non-mergers, with the addition of non-parametric statistics for the KiDS networks. The galaxies identified by these networks are likely to be galaxy mergers that are physically close, either when passing each other or at final coalescence, as visually identified mergers are typically biased towards these merger periods \citep[e.g.][]{2019A&A...626A..49P}. However, that does not exclude the possibility that the network cannot identify galaxy mergers where the merging objects have a greater separation. It is possible to train a CNN to identify merging galaxies that have greater physical separation if such galaxies are present in the training set. However, we do not have a pre-selected sample of such galaxies available as training set for this study and other studies using simulations have shown that larger separation of the merging galaxies can reduce the accuracy of a CNN \citep{2019A&A...626A..49P}.

Details of the architectures of the networks can be found in \citet{2019arXiv190810115P, 2019A&A...626A..49P}. There are 54\,928 mergers (16.1\%) and 30\,033 mergers (29.6\%) identified using CNN in the SDSS sample and the GAMA sample respectively.

In order to demonstrate the difference between the SDSS and KiDS imaging surveys, Figure \ref{cutouts} in the appendix shows cutouts of some example galaxies that are covered by both surveys.

\subsection{AGN classification using MIR color cut and optical emission lines}\label{AGNiden}
We select AGNs in two different ways, one by using a MIR color cut and the other through the BPT diagram. These two methods are combined together to provide a more complete AGN sample. The MIR color cut selection can pick up more dust obsured AGNs that may be missed by optical selection \citep[e.g.,][]{Lacy}.

We cross match our SDSS and GAMA samples with the WISE ALLWISE catalog by selecting the closest pair within a matching radius of 6$\arcsec$, which is close to the angular resolution of 6.1$\arcsec$ in the 3.6 $\mu\text{m}$ band \citep[denoted as W1,][]{Wright}. The angular separation is $< 1\arcsec$ for the vast majority of matched sources (86\% of SDSS sample and 85\% of GAMA sample). We adopt a single color cut $m_{3.6 \mu\text{m}}-m_{4.5 \mu\text{m}}>0.8$ \citep[$W1-W2>0.8$, in Vega magnitudes;][]{Stern12} for $W2<=15$ and require a signal-to-noise ratio S/N>=5 in both bands to select MIR AGNs. We also use the unWISE \citep{Lang} data which provide a new set of coadds of the WISE images that are not blurred, and try a $W1-W2>0.5$ \citep{Assef13} criterion for both sets of WISE data, since \citet{Blecha} argue that the $W1-W2>0.5$ cut can greatly improve completeness without significantly decreasing reliability. 
However, since the results are very similar using these two catalogs, and the two color cuts, for the analysis presented here we only show the $W1-W2>0.8$ AGN selection from ALLWISE.

Besides the MIR AGNs, we also use optical emission line diagnostics to classify optical AGNs. For the SDSS sample, we use the MPA-JHU spectroscopic analysis to obtain BPT AGN classification based on emission line ratios. This classification follows procedures described in \citet{Brinchman} which used demarcation lines from \citet{Kewley01b} to ensure a minimum contamination to the fluxes from star formation. We also adopt the \citet{Schawinski07} criteria to exclude Low Ionization Nuclear Emission Line Regions (LINERs), since whether LINERs can be recognized as AGNs is still debated \citep{Maoz, Yan12, Singh}. 

Optically selected AGNs in the GAMA sample at $z<0.3$ (for a reliable detection of H$\alpha$ line) are obtained using emission line information from the SpecLineSFRv05 catalog \citep{Gordon17}. Specifically, narrow-line AGNs are selected by requiring BPT diagnostics satisfying both the \citet{Kewley01b} and \citet{Schawinski07} criteria for the Seyfert classification, excluding LINERs. Where either H$\beta$ or [OIII]5007$\AA$ lines are not detected, AGN classification is done via WHAN diagnostics \citep[W$_{H\alpha}$ vs N II/H$\alpha$,][]{Cid} using the criteria of \citet{Gordon18}.

We refer to AGNs selected by the WISE color cut and optical emission line information as MIR AGN and OPT AGN respectively. There are 421 and 96 AGNs that are both OPT and MIR AGNs in the SDSS and GAMA sample respectively. We do not split MIR or OPT AGNs further into subgroups according to whether they are also classified in the other method, in order to have enough number of MIR and OPT AGNs in the analysis below.

For MIR AGNs, we use the rest-frame $6\,\mu\text{m}$ luminosity to trace the AGN accretion power. Continuum emission at this wavelength is believed to originate from the dusty torus that absorbs ultra-violet/optical photons from the accretion disk and then re-emits at longer wavelengths \citep{Lutz04, Gandhi, Mateos15}. The rest-frame $6\,\mu\text{m}$ flux is derived by linearly interpolating the WISE W1, W2, W3 ($12\,\mu\text{m}$) bands fluxes (after shifting to rest-frame). 

For OPT AGNs we use the [O III] 5007$\AA$ line luminosity as an indicator of the AGN accretion power. According to the unification model of AGN \citep{UP95, Antonucci}, [O III] is radiated by gas in the narrow line region (NLR) which is located outside of the torus, hence experiences moderate dust obscuration \citep{Kauffmann03, Heckman05}. [O III] luminosity is corrected for extinction using the Balmer decrement (assuming an intrinsic value of 3.0) according to the following equation \citep{Bassani, Lamastra} where $L_{O III}^{c}$ is the extinction-corrected [O III] luminosity:
\begin{equation*}
L_{OIII}^{c}=L_{OIII}\left(\frac{(\text{H}\alpha/\text{H}\beta)_{obs}}{3.0}\right)^{2.94} \tag{4}
\label{eq4}
\end{equation*}
We also use the $L_{6\,\mu\text{m}}-L_{2-10\,\text{keV}}$ relation from \citet{Mateos15} and the $L_{[O III]}-L_{2-10\,\text{keV}}$ relation from \citet{Heckman05} to transform the rest-frame $6\,\mu\text{m}$ luminosity and [O III] line luminosity into X-ray luminosities, serving as a common proxy of the bolometric power for both MIR and OPT AGNs. Due to the large scatter of these relations ($\sim$ 0.4 dex and 0.5 dex respectively), we only use $L_{2-10\,\text{keV}}$ as a rough indicator of AGN bolometric power. 

\begin{figure*}
\resizebox{\hsize}{!}{\includegraphics[width=\linewidth]{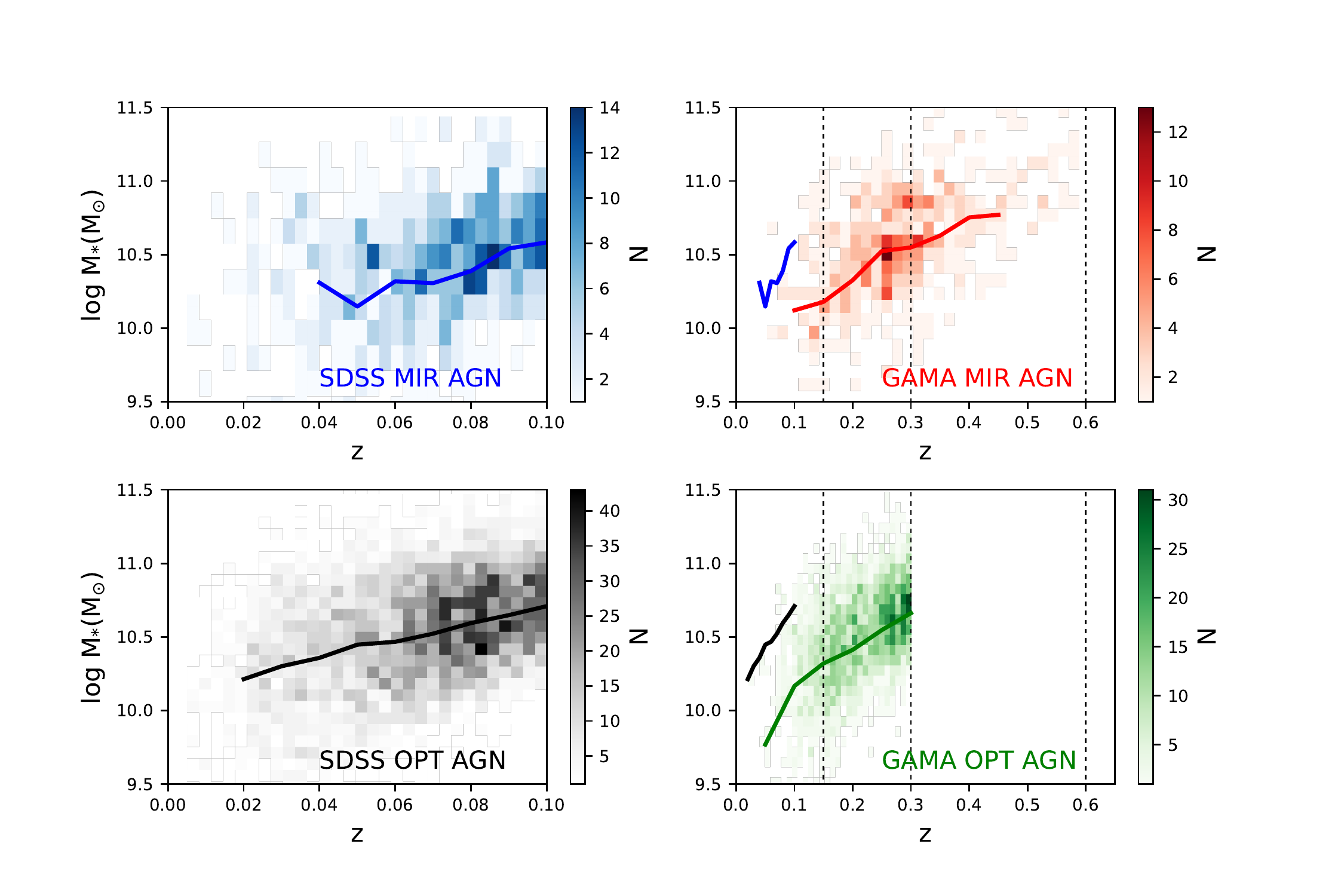}}
\caption{Stellar mass M$_{*}$ vs redshift $z$ distributions for the SDSS and GAMA AGNs. Red and green distributions represent MIR and OPT AGNs for the GAMA sample respectively. Blue and grey distributions represent MIR and OPT AGNs for the SDSS sample respectively. The solid lines indicate the running median for each group. The dashed lines mark the edges of the redshift bins for the GAMA AGNs (see Section \ref{r1}). In the SDSS sample, the OPT AGNs are hosted in more massive galaxies than the MIR AGNs. In the same redshift range, the GAMA AGNs are hosted in less massive galaxies than the SDSS AGNs. }
\label{distribution1}
\end{figure*}

\begin{figure*}
\resizebox{\hsize}{!}{\includegraphics[width=\linewidth]{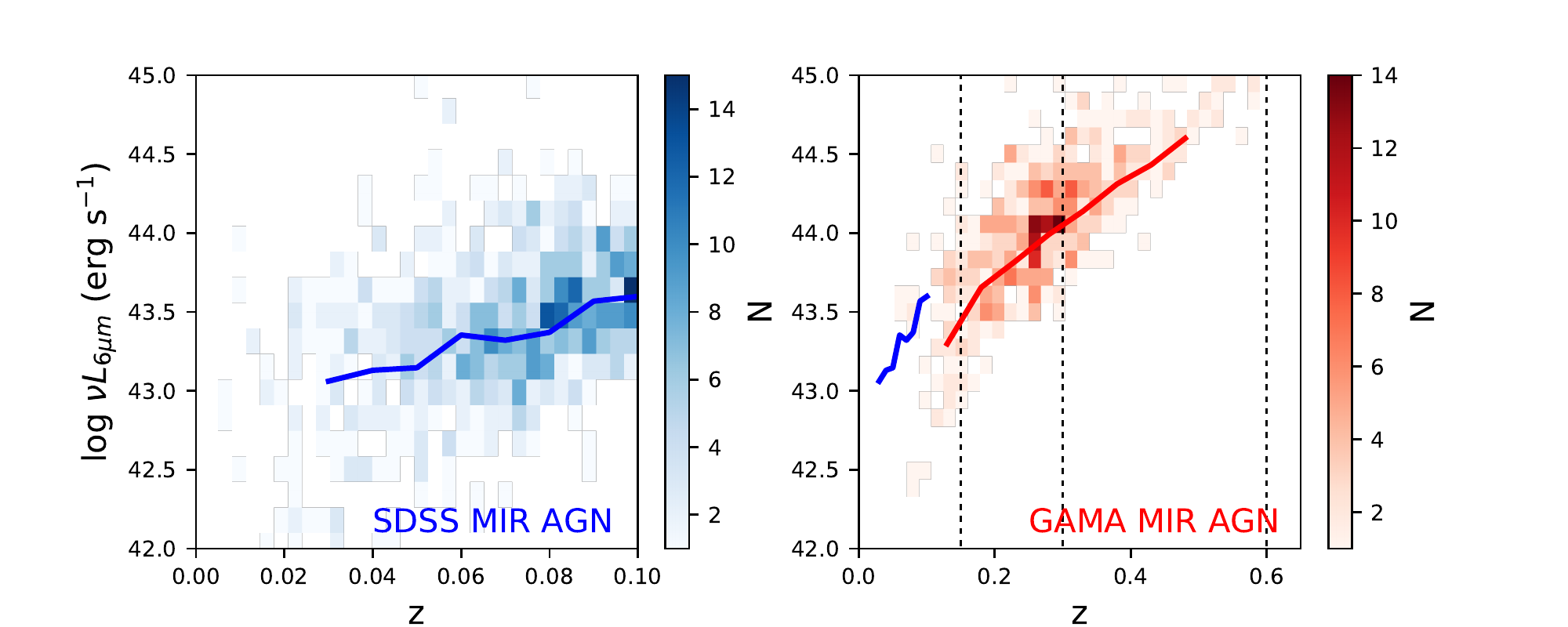}}
\caption{Rest-frame $6\,\mu\text{m}$ luminosity $L_{6\mu m}$ vs redshift $z$ distributions for the SDSS and GAMA MIR AGNs. The solid lines indicate the running median for each group. The dashed lines mark the edges of the redshift bins for the GAMA MIR AGNs. In the same redshift range, the GAMA MIR AGNs are less powerful than the SDSS MIR AGNs.} 
\label{distribution2}
\end{figure*}

\begin{figure*}
\includegraphics[width=\linewidth]{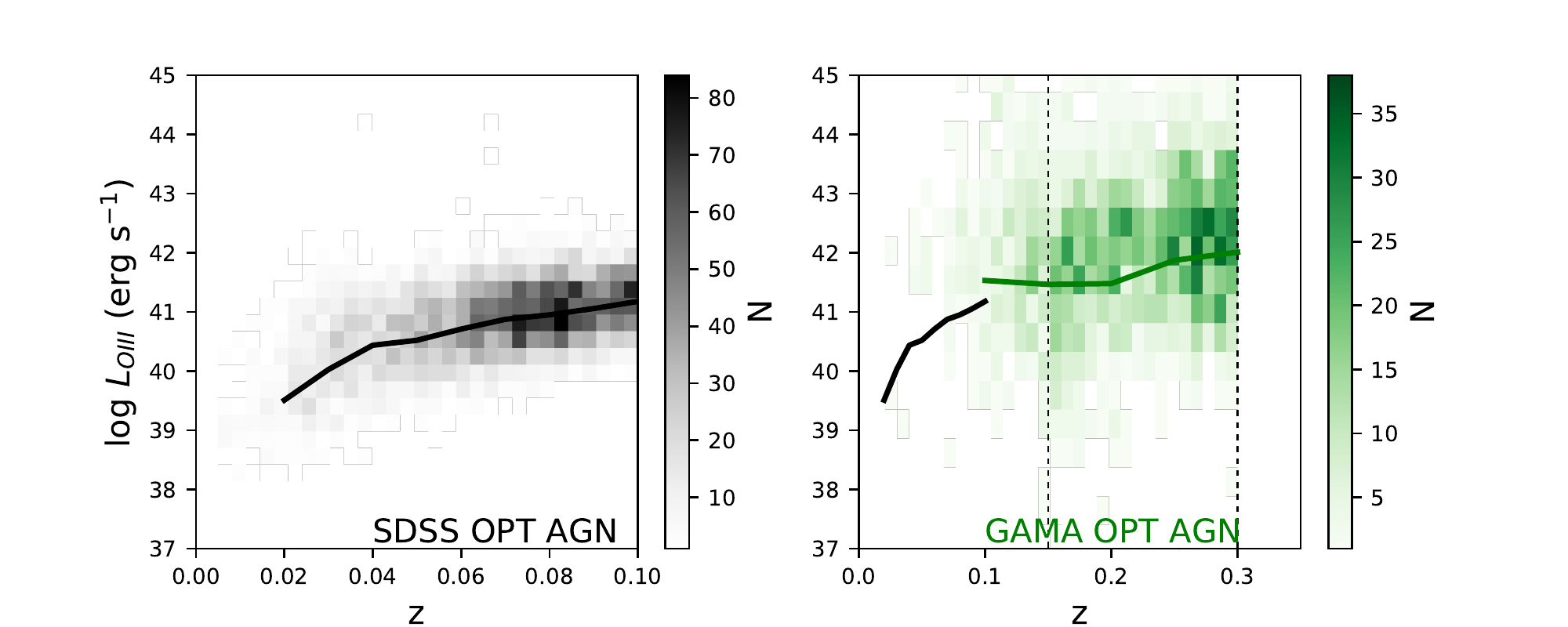}
\caption{[O III] luminosity $L_{OIII}$ vs redshift $z$ distributions for the SDSS and GAMA OPT AGNs. The solid lines indicate the running median for each group. The dashed lines mark the edges of the redshift bins for the GAMA OPT AGNs. In the same redshift range, the GAMA OPT AGNs are less powerful than the SDSS OPT AGNs.}
\label{distribution3}
\end{figure*}

\begin{figure*}
\centering

 \begin{subfigure}{0.4\textwidth}
  \includegraphics[width=\linewidth]{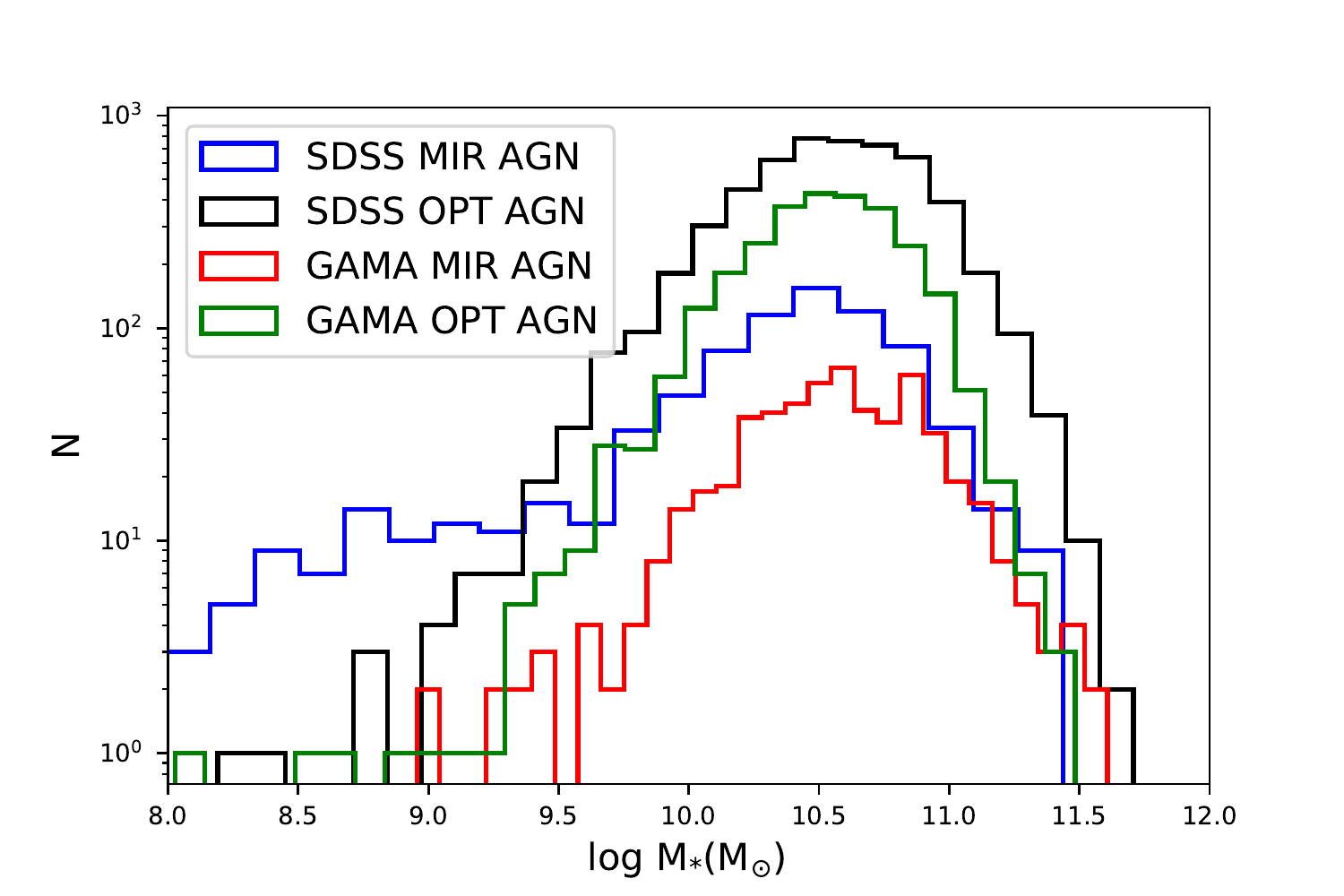}
 \end{subfigure}
 \begin{subfigure}{0.4\textwidth}
  \includegraphics[width=\linewidth]{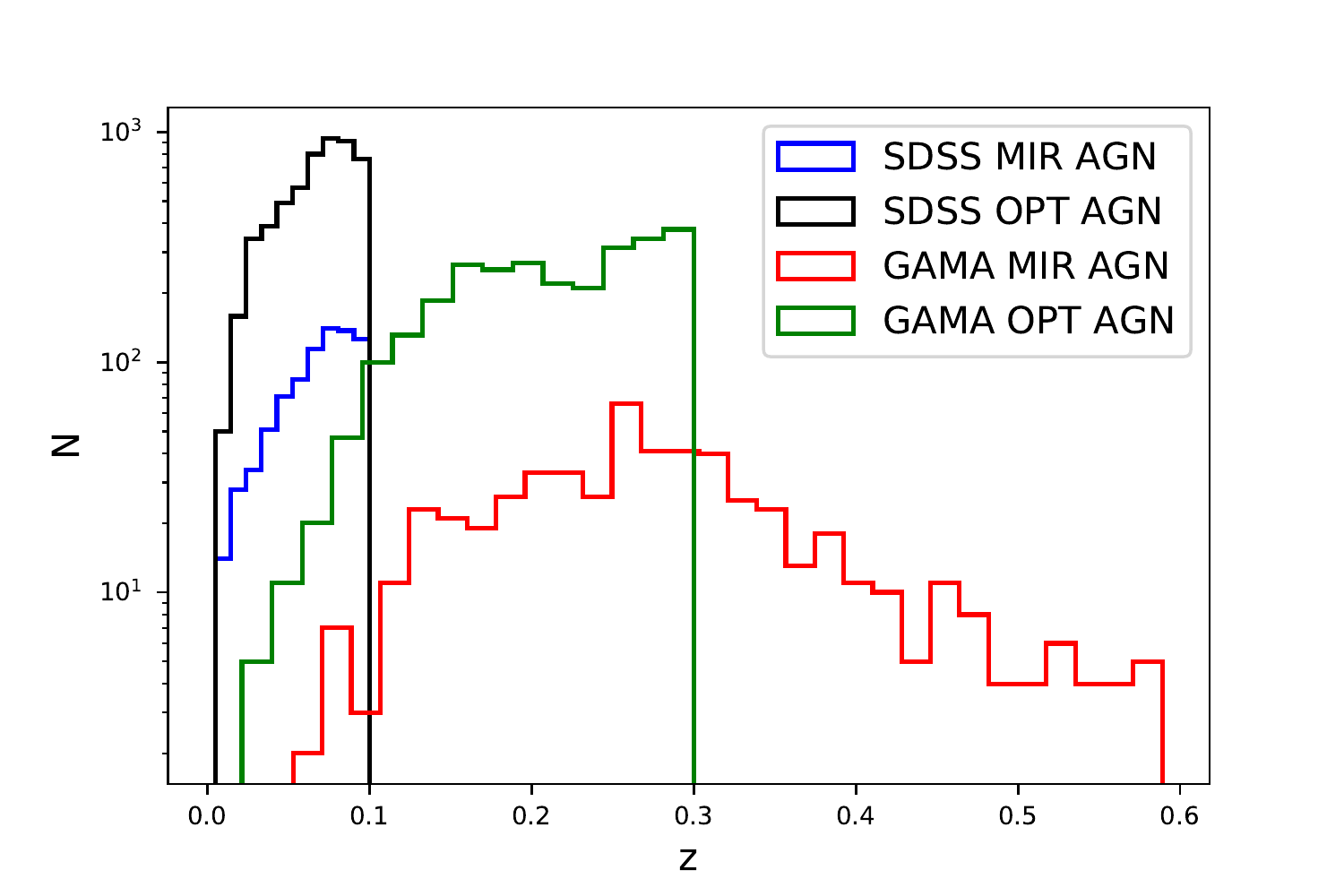}
 \end{subfigure}
 \begin{subfigure}{0.4\textwidth}
  \includegraphics[width=\linewidth]{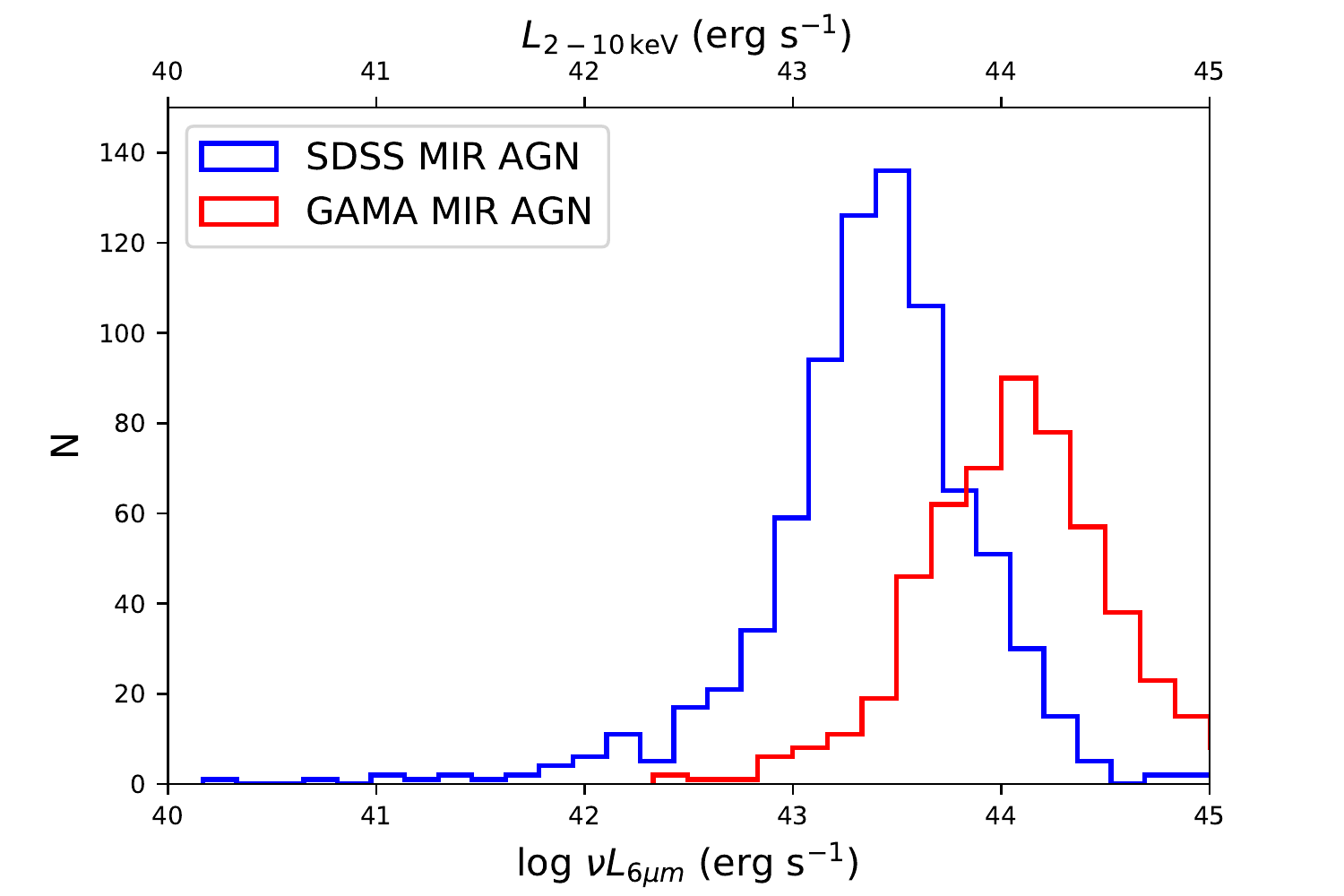}
 \end{subfigure}
 \begin{subfigure}{0.4\textwidth}
  \includegraphics[width=\linewidth]{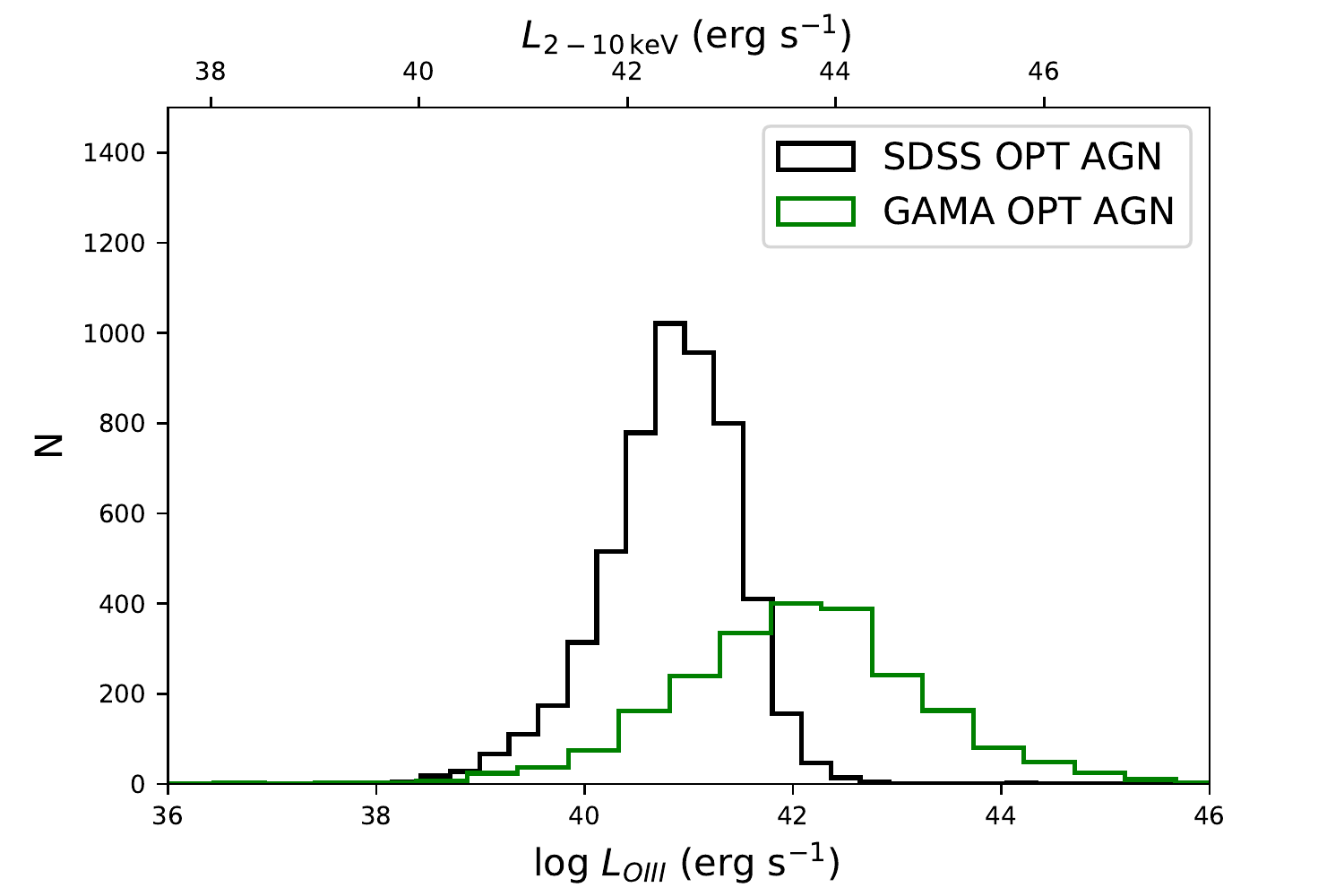}
 \end{subfigure}
\caption{Top: Mass and redshift distributions of the MIR and OPT AGNs in the SDSS and GAMA samples. Bottom: Rest-frame $6\,\mu\text{m}$ ([O III]) luminosity distribution of the MIR (OPT) AGNs in the SDSS and GAMA samples.}

\label{distribution4}
\end{figure*}

Figures \ref{distribution1}, \ref{distribution2} and \ref{distribution3} show the distributions of the SDSS and GAMA AGNs in the M$_{*}-z$, $L_{6\,\mu\text{m}}-z$ and $L_{\rm [O III]}-z$ parameter space respectively. Figure \ref{distribution4} shows the histograms of each parameter. In Figure \ref{distribution1} we can see that the host galaxies of SDSS OPT AGNs are more massive than the host galaxies of SDSS MIR AGNs. This is also clear from the histogram of stellar mass distribution in panel (a) of Figure \ref{distribution4}. In addition, the host galaxies of the MIR and OPT AGNs in the GAMA sample are less massive than the host galaxies of the MIR and OPT AGNs in the SDSS sample in the same redshift range. Similarly in Figure \ref{distribution2} and \ref{distribution3}, the MIR and OPT AGNs in the GAMA sample are less powerful than the MIR and OPT AGNs in the SDSS sample in the same redshift range. 

\subsection{Low accretion rate AGNs}

\begin{figure*}
\resizebox{\hsize}{!}{\includegraphics[]{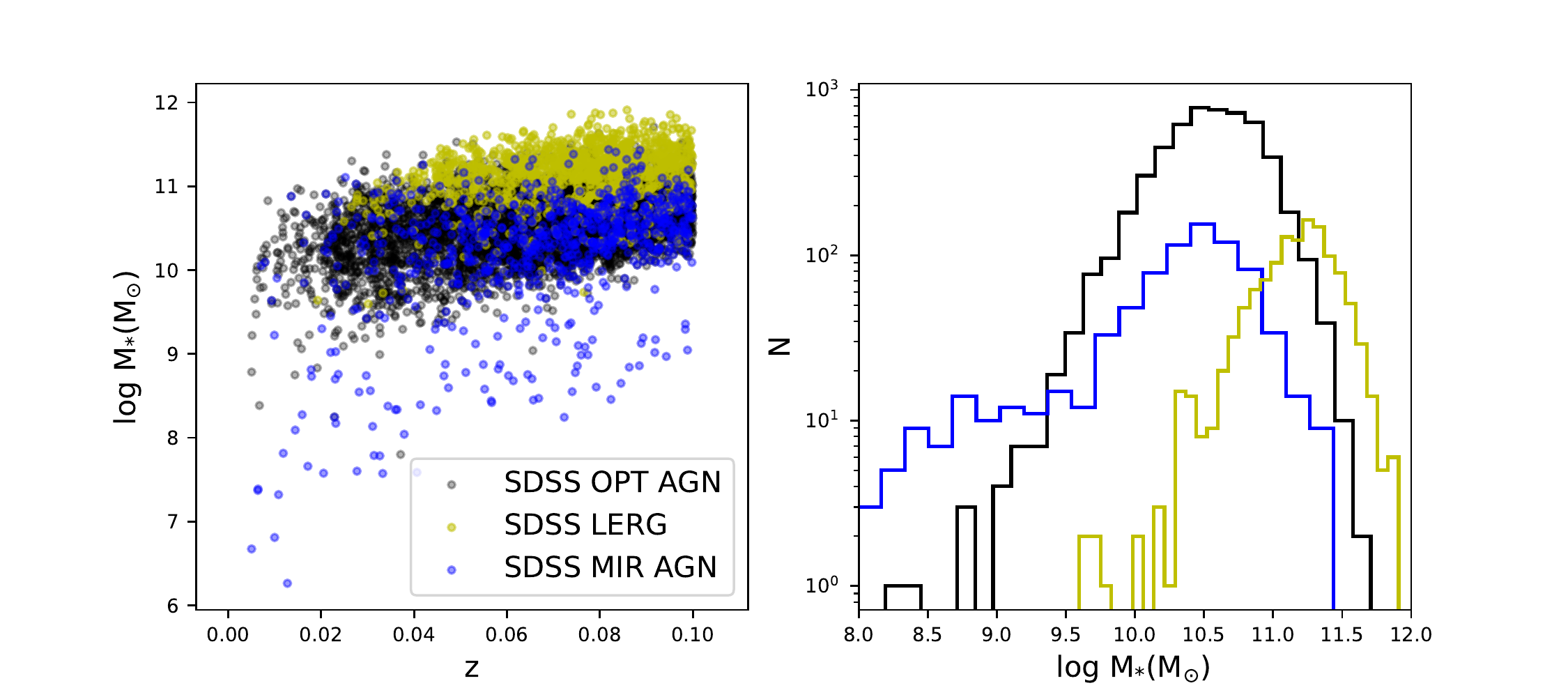}}

\caption{Left: Mass and redshift distributions of the MIR AGNs, OPT AGNs and LERGs in the SDSS sample. Right: Histograms of mass distributions. We can see that LERGs are more massive.}
\label{lergcomp}
\end{figure*}

Studies of radio AGNs divide them into two distinct types according to the mode of feedback, one with strong radiation (high-excitation radio galaxies, HERGs) and the other with jets \citep[low-excitation radio galaxies, LERGs;][]{BH12}. These two types are believed to have different black hole accretion rates, with HERGs accreting at a higher rate than LERGs \citep[e.g.,][]{smolcic, Janssen, Mingo}. Previous studies propose that highly accreting AGNs are triggered by large gas reservoirs brought into the central SMBHs by mergers while low accreting rate AGNs are fueled by smaller amount of gas transported by secular processes \citep{Heckman86, BH12, Tadhunter}. These results are supported by a normal LERG fraction in galaxy mergers \citep{Ellison15}, with the exception of the low mass merger population \citep{Gordon19}. We select LERGs from \citet{BH12} matched with our SDSS merger identification and build a control sample, totaling 1225 LERGs and 11\,250 mass and redshift matched non-LERG controls, in order to investigate the connection between mergers and these low accretion rate AGNs. There are only 5 (20) LERGs that are also MIR (OPT) AGNs, suggesting that they may represent a different evolutionary stage of AGNs. Figure \ref{lergcomp} shows mass and redshift distributions of the MIR AGNs, OPT AGNs and LERGs in the SDSS sample. It is clear that the LERGs are in general more massive.

\section{Results}\label{re}

In this work, we investigate the AGN-merger connection from two angles. We first study the AGN fractions in mergers/non-mergers to assess whether mergers are a viable triggering mechanism. In this first experiment, the signature that mergers are able to trigger AGN is a higher fraction of AGN in the mergers sample, compared to the non-mergers. To address the first aspect, we start from a merger sample and a non-merger control sample, and investigate the difference in the AGN fraction in these two samples. 

In the second experiment we study the merger fractions in AGN/non-AGNs in order to find out whether mergers dominate the triggering of AGNs \citep[also see][]{Ellison19}. To address the second aspect, we start from a AGN sample and a non-AGN control sample, for the MIR as well as the optical AGN selection methods, and investigate the merger fraction in these two samples.

Following \citet{Ellison11}, for each merger in the SDSS and GAMA samples, we identify a non-merger counterpart satisfying the following requirements. 

\begin{equation*}
|z_{\rm control}-z_{\rm sample}| \leq 0.01  \tag{1}
\label{eq2}
\end{equation*}
\begin{equation*}
\rm |log\,M_{*}^{control}-log\,M_{*}^{sample}| \leq 0.1 \text{dex} \tag{2}
\label{eq3}
\end{equation*}

For the first experiment we only include mergers that have no fewer than 10 non-merger counterparts and randomly choose 10 of them to establish a non-merger control sample. For the second experiment we first adopt a conservative method for building the non-AGN control samples. For the MIR AGNs we set $W1-W2<0.5$ when selecting controls and for OPT AGNs we exclude composites when selecting controls, in order to ensure a minimum AGN contamination in the control samples. Similar to the first experiment, we then randomly select 10 non-AGN counterparts for each AGN satisfying the above requirements and exclude AGNs that have fewer than 10 counterparts. The non-AGN control samples for MIR AGNs do not contain OPT AGNs and vice versa. We note here that the sources in the control group are not necessarily unique, and some of them may appear more than once. Table \ref{tb1} shows the number of galaxies in each subgroup. The number of the galaxies in the control sample for each subgroup is 10 times larger.

Besides the main merger sample described above, we also build a stricter merger sample and non-merger control sample. The output of the CNN can be seen as the a probability for a galaxy to be a merger. We increase the threshold for identifying mergers and decrease the threshold for selecting non-mergers, in both SDSS and GAMA samples. For example, the SDSS CNN uses $p_{\text{merger}}>0.57$ as merger threshold. We raise this threshold to $p_{\text{merger}}>0.8$ for a stricter and less contaminated merger sample and lower this threshold to $p_{\text{merger}}<0.4$ for selecting more conservative non-merger controls.

\begin{table*}
\caption{Number of sources in each group. The size of sources in the control sample for each subgroup is 10 times larger. The main merger sample is the default merger sample in our study. We apply a stricter merger selection to build a stricter merger sample.}
\label{tb1}      
\centering
\begin{tabular}{c c c c c c }
\hline\hline
survey&all&merger\_main&merger\_stricter&MIR AGN&OPT AGN\\
\hline
SDSS&341\,908&54\,642&33\,151&799&5420\\
$0.005<z<0.1$&&&&&\\
\hline
GAMA&101\,470&29\,560&19\,231&543&2750\\
$0<z<0.6$&&&&&0<z<0.3\\
\hline
\end{tabular}
\end{table*}

\subsection{AGN fractions in mergers and non-mergers}\label{r1}
\begin{figure*}
\begin{subfigure}{.5\textwidth}
\centering
\includegraphics[width=\linewidth]{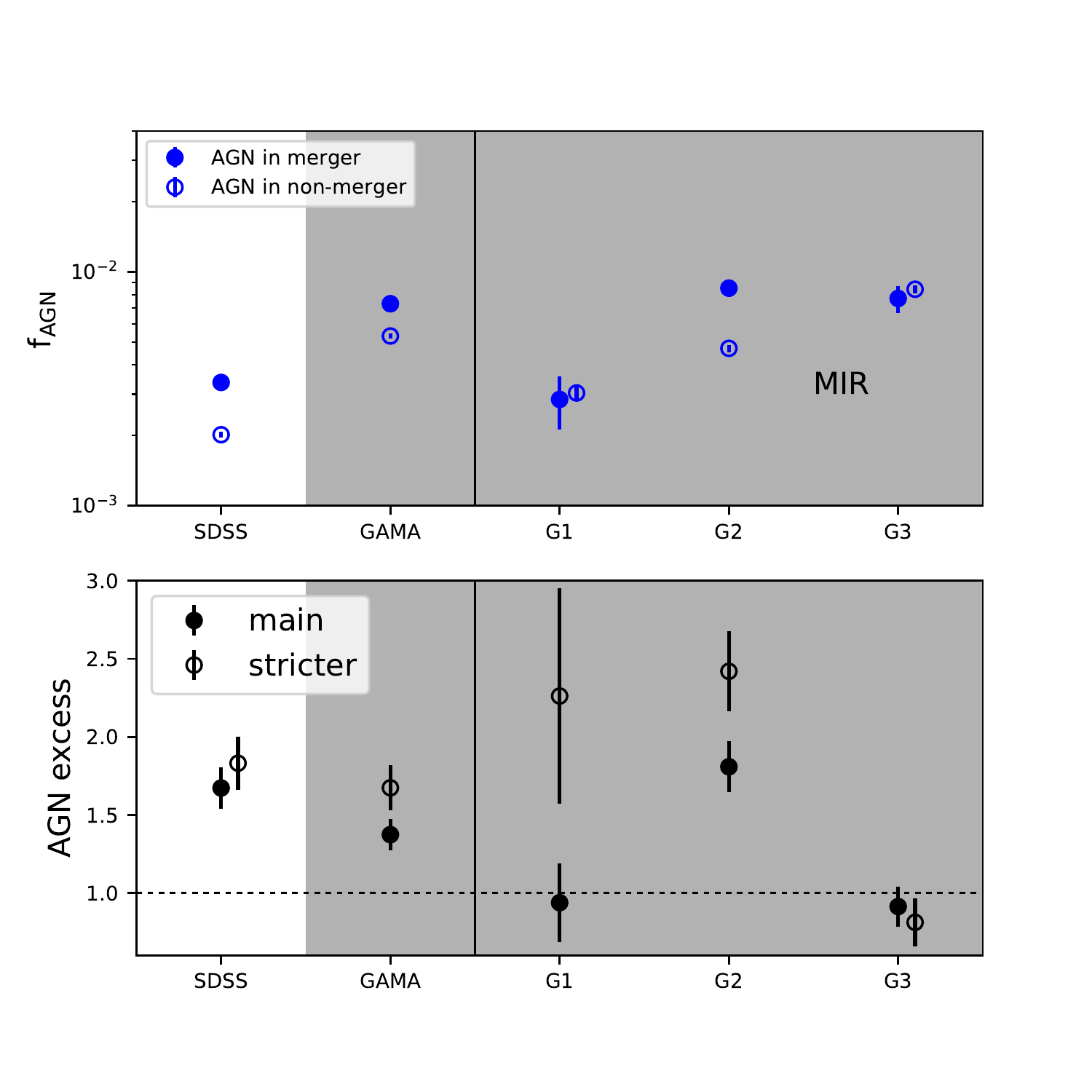}
\end{subfigure}
\begin{subfigure}{.5\textwidth}
\centering
\includegraphics[width=\linewidth]{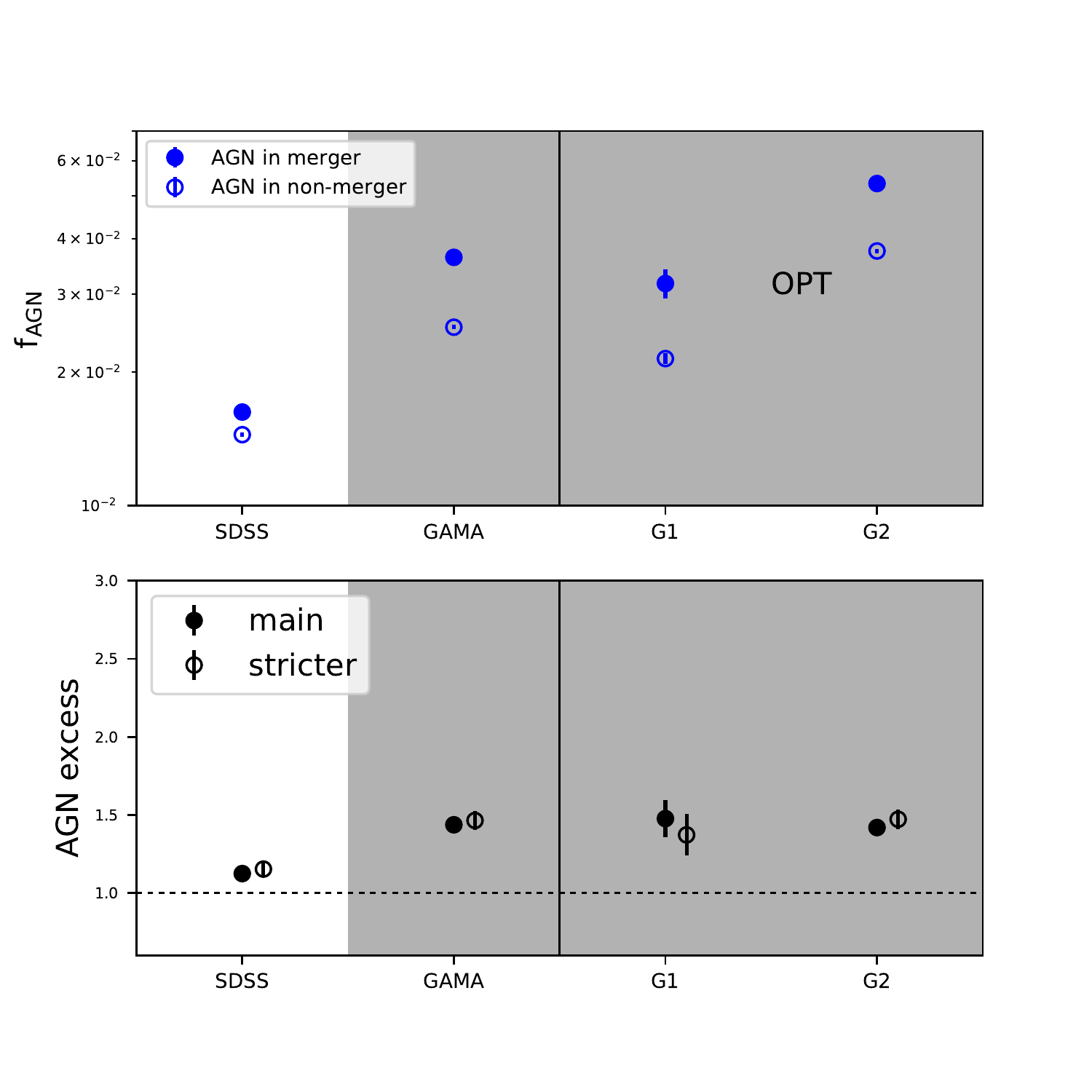}
\end{subfigure}
\caption{Left: MIR AGN fractions in mergers/non-mergers for the SDSS sample (white background) and the GAMA sample (shadowed background). We also divide the GAMA sample into three redshift bins (denoted as G1, G2, G3). Right: OPT AGN fractions in mergers/non-mergers for the SDSS and GAMA sample. We also divide the GAMA sample into two redshift bins. Errors are calculated through binomial statistics. In the bottom panels we plot the ratio of AGN fraction in mergers relative to that in non-mergers (i.e., AGN excess) for the main merger sample and the stricter merger sample. The dashed lines indicate the excess value of one which means no difference in the AGN fraction in mergers relative to non-mergers. We find a qualitatively consistent picture between the SDSS and GAMA samples in which the AGN fraction is higher in mergers compared to non-mergers. }
\label{agn_in_merger}
\end{figure*}

First, we focus on the AGN fraction in mergers and matched non-merger galaxies to explore whether mergers can trigger AGN. 
Figure \ref{agn_in_merger} shows the comparisons of AGN fractions in mergers/non-mergers based on the MIR and optical selections. According to the merger classification networks (see Section \ref{merger}), we also separate the GAMA sample into three redshift bins: $0<z<0.15$ (denoted as G1), $0.15<z<0.3$ (denoted as G2) and combine $0.3<z<0.45$ and $0.45<z<0.6$ into one redshift bin (denoted as G3) in the right half of each panel. G3 is not included in the GAMA OPT AGNs because the optical emission line diagnosis in the GAMA sample is limited to $z<0.3$ since the H$\alpha$ line at higher redshift will move outside of the spectral range of the spectrograph used in the GAMA survey \citep[see][]{Gordon17}.

\begin{table*}
\caption{The MIR and OPT AGN fractions in mergers/non-mergers for the SDSS sample and the GAMA (sub)samples. Errors are calculated through binomial statistics.}
\label{tb2}      
\centering
\begin{tabular}{c c c c c}
\hline\hline

&MIR AGN&MIR AGN&OPT AGN&OPT AGN\\
&in merger&in non-merger&in merger&in non-merger\\
\hline
SDSS&$0.34\pm{0.02}\%$&$0.20\pm{0.01}\%$&$1.63\pm{0.05}\%$&$1.45\pm{0.02}\%$\\
&(184/54642)&(1100/546420)&(889/54642)&(7905/546420)\\
\hline
GAMA&$0.73\pm{0.05}\%$&$0.53\pm{0.01}\%$&$3.63\pm{0.11}\%$&$2.53\pm{0.03}\%$\\
&(216/29560)&(1572/295600)&(1074/29560)&(7473/295600)\\
\hline
G1&$0.28\pm{0.07}\%$&$0.30\pm{0.02}\%$&$3.17\pm{0.24}\%$&$2.15\pm{0.06}\%$\\
$0<z<0.15$&(15/5266)&(164/54019)&(167/5266)&(1160/54019)\\
\hline
G2&$0.85\pm{0.07}\%$&$0.47\pm{0.02}\%$&$5.33\pm{0.17}\%$&$3.75\pm{0.05}\%$\\
$0.15<z<0.3$&(145/17018)&(792/168124)&(907/17018)&(6313/168124)\\
\hline
G3&$0.77\pm{0.10}\%$&$0.84\pm{0.03}\%$&--&--\\
$0.3<z<0.6$&(56/7276)&(619/73457)&--&--\\
\hline
\end{tabular}
\end{table*}

We find that in general the AGN fraction in mergers is larger than that in non-mergers for both the SDSS and GAMA samples and both AGN selection methods. $1.63\pm{0.05}\%$ and $0.34\pm{0.02}\%$ of mergers in the SDSS sample host OPT AGNs and MIR AGNs respectively, while the percentages of non-mergers in the SDSS sample are $1.45\pm{0.02}\%$ and $0.20\pm{0.01}\%$ respectively.  $3.63\pm{0.11}\%$ and $0.73\pm{0.05}\%$ of mergers in the GAMA sample host OPT AGNs and MIR AGNs respectively, while the percentages of non-mergers in the GAMA sample is $2.53\pm{0.03}\%$ and $0.53\pm{0.01}\%$. Although the overall AGN fraction is low, we can still see a slight enhancement of AGN fraction in mergers than in non-mergers (up to $\sim 1.5$ AGN excess), suggesting that mergers do trigger AGN activity. When applying a stricter merger sample, generally we observe a greater AGN excess, which also supports our argument. In addition, the MIR AGN excess is larger than OPT AGN excess when a less contaminated merger sample is applied, which may imply that mergers are more important in triggering MIR AGNs. The numbers of galaxies and fractions for each group are listed in Table \ref{tb2}.

Although it seems that the lowest redshift ($0<z<0.15$) and the highest redshift ($0.3<z<0.45$) bin of GAMA MIR AGNs shows an inverse trend with more MIR AGNs found in non-mergers, we argue that they are not significant (within $1\sigma$ uncertainty). Comparing horizontally between the SDSS sample and the GAMA sample in the lowest redshift bin in the two panels in Figure \ref{agn_in_merger}, it seems that for the MIR AGNs, the SDSS sample agrees well with the GAMA sample in the lowest redshift bin, while for the OPT AGNs, the SDSS sample shows a lower AGN fraction in mergers and non-mergers. We note here that unlike the MIR AGN classification, the OPT AGN classification in the GAMA sample adopted by \citet{Gordon17} is slightly different from that in the SDSS sample, including OPT AGNs that may be missed when H$\beta$ or [OIII]5007$\AA$ lines are not detected (see Section \ref{AGNiden}).


Compared to previous studies, we find a smaller contrast of AGN fractions in mergers and non-mergers, with an AGN excess of up to $\sim1.5$ in mergers relative to non-merger controls. \citet{Ellison11} found an increase of AGN fraction by a factor of 2.5 in galaxy pairs relative to control sample, using a sample of 11\,060 SDSS pairs. \citet{Silverman11} found that galaxy pairs are 1.9 times more likely to host X-ray AGNs than mass-matched isolated galaxies, for a sample of 562 galaxies in pairs at $0.25<z<1.05$. \citet{Satyapal} found that as the separation between galaxies decreases, the excess of MIR AGN fraction in pairs relative to controls increases, reaching a factor of 10-20 of AGN excess in post-mergers. \citet{Weston} found it 5-17 (3-5) times more likely for mergers to host MIR selected AGNs compared to non-mergers, for a sample of 130 mergers (1069 interactions) with stellar mass above $2 \times 10^{10}\, \text{M}_{\odot}$. \citet{Goulding} found that mergers are 2-7 times more likely to host obscured MIR AGNs than non-interacting galaxies with a sample of 2552 obscured AGNs. 

Even though our work adopts a different method (CNN) to identify mergers than visual inspection, a different definition of mergers \citep[e.g., galaxy pairs in][]{Ellison11, Satyapal}, a slightly different threshold in classifying AGNs, as well as different sample distributions in terms of redshift and stellar mass, our results are qualitatively consistent with previous studies. In addition, when applying a stricter merger selection which is less affected by contamination, the AGN excess can be more than 2, which is more consistent with previous studies.

\subsection{Merger fractions in AGNs and non-AGNs}\label{r2}
\begin{figure*}
\begin{subfigure}{.5\textwidth}
\centering
\includegraphics[width=\linewidth]{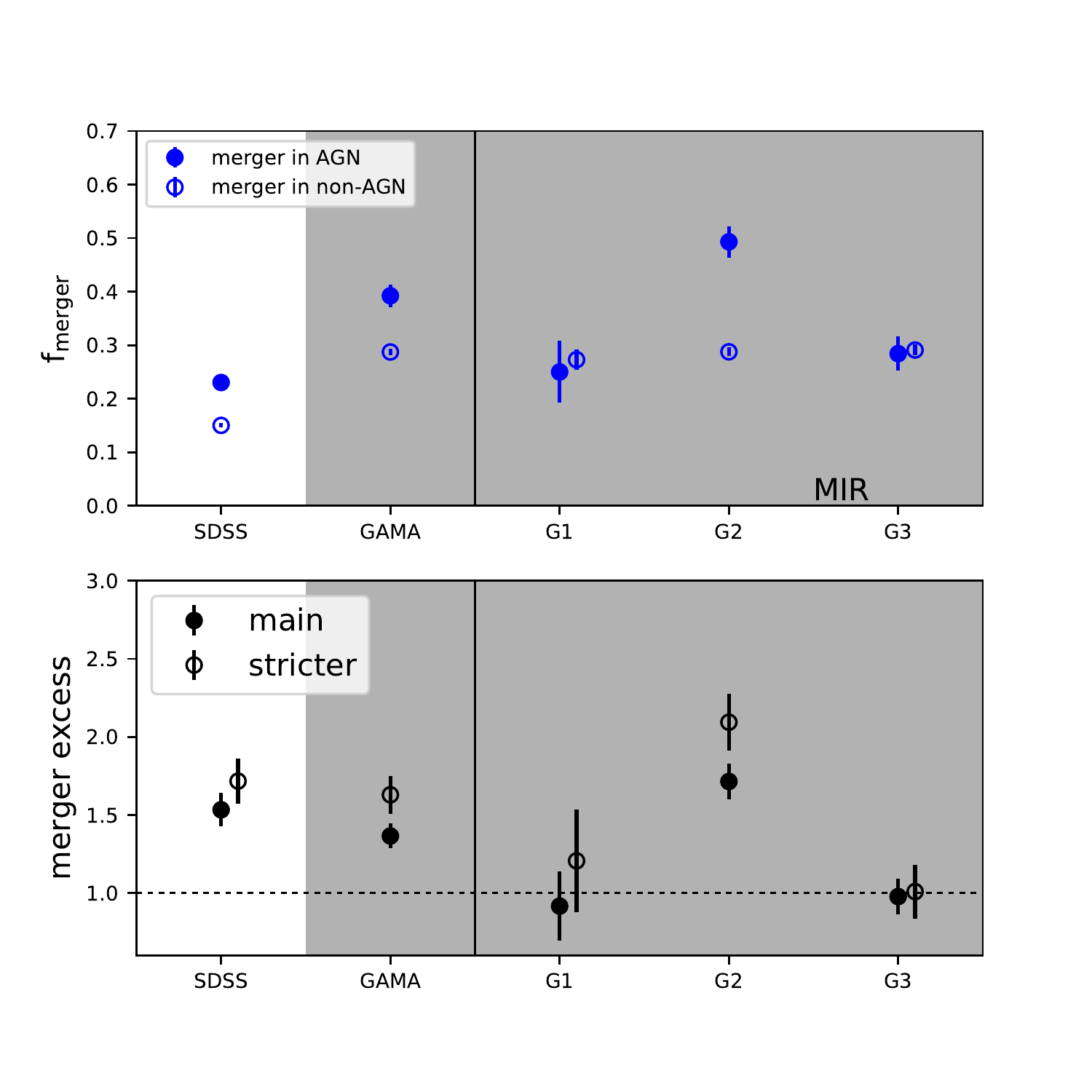}
\end{subfigure}
\begin{subfigure}{.5\textwidth}
\centering
\includegraphics[width=\linewidth]{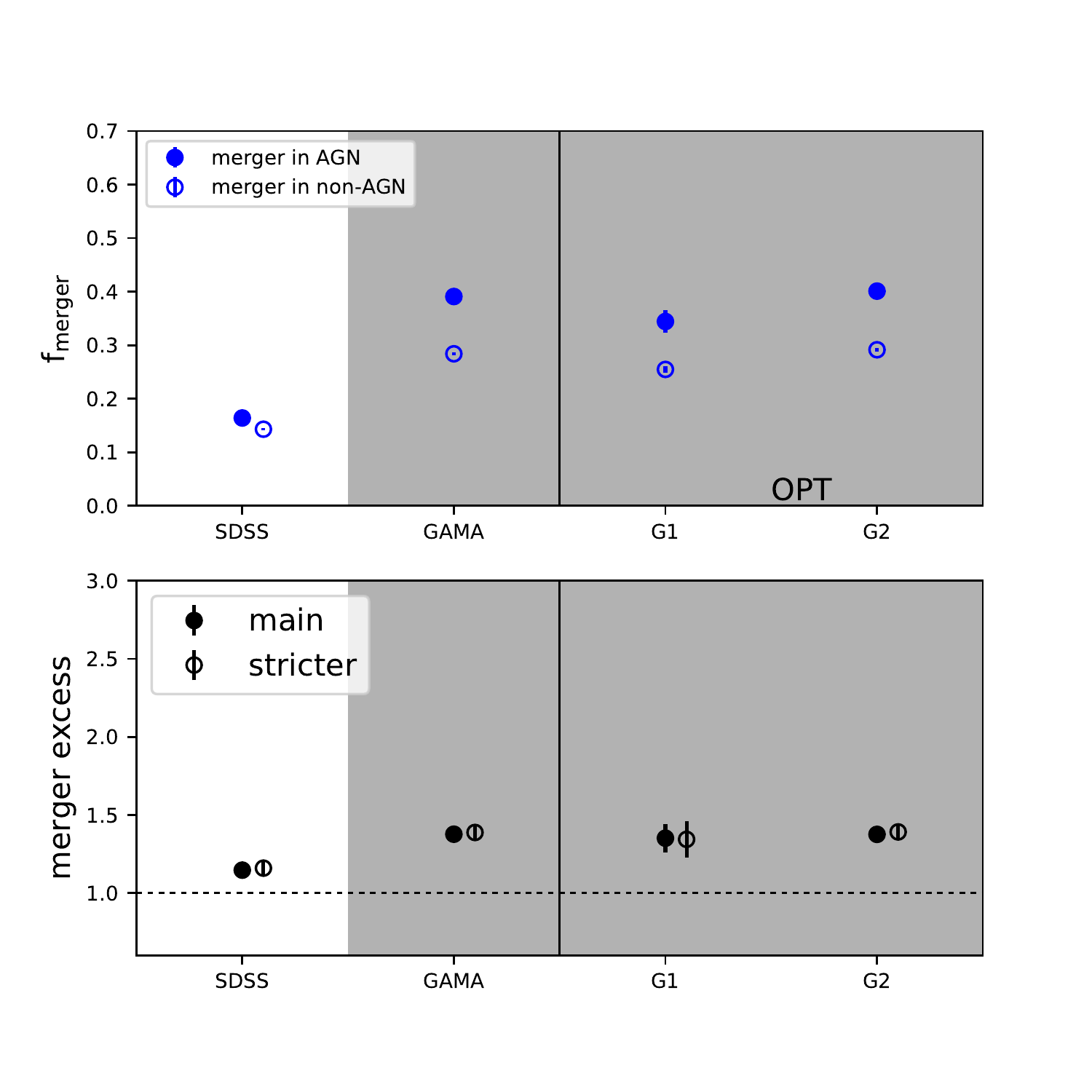}
\end{subfigure}
\caption{Left: Merger fractions in MIR AGNs/non-AGNs for the SDSS and GAMA sample. We also divide the GAMA sample into three redshift bins.  Right: Merger fractions in OPT AGNs/non-AGNs for SDSS and GAMA sample. We also divide the GAMA sample into two redshift bins. Errors are calculated through binomial statistics. In the bottom panels we plot the ratio of merger fraction in AGNs relative to that in non-AGNs (i.e., merger excess) for the main sample and a stricter merger identification threshold. The dashed lines indicate the excess value of one which means no difference in the merger fraction in AGNs relative to non-AGNs. Again, we find a qualitatively consistent picture between the SDSS and GAMA samples in which the merger fraction is higher in AGNs compared to non-AGNs.}
\label{merger_in_AGN}
\end{figure*}

In Section \ref{r1} we assess whether mergers exhibit an enhanced AGN fraction. In this section, we perform the reverse experiment, by assessing whether AGNs are preferentially hosted by merging galaxies compared to non-AGN controls.
The left and right panels of Figure \ref{merger_in_AGN} show the comparisons of merger fractions in AGN/non-AGNs based on MIR and optical selections respectively. We also separate the GAMA sample into different redshift bins in the right half of each panel.

\begin{table*}
\caption{The merger fractions in AGNs/non-AGNs for the SDSS and GAMA sample. Errors are calculated through binomial statistics.}
\label{tb3}      
\centering
\begin{tabular}{c c c c c}
\hline\hline

&merger in&merger in&merger in&merger in\\
&MIR AGN&MIR control&OPT AGN&OPT control\\
\hline
SDSS&$23.03\pm{1.49}\%$&$15.02\pm{0.40}\%$&$16.40\pm{0.50}\%$&$14.30\pm{0.15}\%$\\
&(184/799)&(1200/7990)&(889/5420)&(7753/54200)\\
\hline
GAMA&$39.23\pm{2.10}\%$&$28.73\pm{0.61}\%$&$39.09\pm{0.93}\%$&$28.39\pm{0.27}\%$\\
&(213/543)&(1560/5430)&(1075/2750)&(7808/27500)\\
\hline
G1&$25.00\pm{5.79}\%$&$27.29\pm{1.91}\%$&$34.43\pm{2.15}\%$&$25.48\pm{0.63}\%$\\
$0<z<0.15$&(14/56)&(149/546)&(168/488)&(1219/4785)\\
\hline
G2&$49.31\pm{2.94}\%$&$28.76\pm{0.83}\%$&$40.10\pm{1.03}\%$&$29.15\pm{0.30}\%$\\
$0.15<z<0.3$&(143/290)&(849/2952)&(907/2262)&(6501/22305)\\
\hline
G3&$28.43\pm{3.21}\%$&$29.09\pm{1.03}\%$&--&--\\
$0.3<z<0.6$&(56/197)&(562/1932)&--&--\\
\hline
\end{tabular}
\end{table*}

From Figure \ref{merger_in_AGN} we can see that the fraction of AGNs that are mergers is higher than the fraction of non-AGN controls that are mergers, for both samples and both AGN selections. More than 16\% of MIR and OPT AGNs in the SDSS sample are merging and $\sim 40\%$ of MIR and OPT AGNs in the GAMA sample are merging, while the fraction of merging control galaxies is $\sim 15-29\%$. Our findings are qualitatively in agreement with previous studies in which AGNs show higher merger fraction than non-AGN control sample, \citet[e.g.,][]{Hong, Ellison19}. If we limit our mergers to the less contaminated ones (with stricter threshold), we can observe an increase of the merger excess in the AGNs relative to non-AGNs. Similar to Figure \ref{agn_in_merger}, the merger excess in MIR AGNs is larger than that in OPT AGNs when a stricter merger threshold is applied, suggesting that mergers may be more important in MIR AGN triggering \citep[e.g., see][]{Ellison19}. Although it seems that the lowest redshift ($0<z<0.15$) and the highest redshift ($0.3<z<0.45$) bin of GAMA MIR AGNs shows an inverse trend with more mergers  found in non-AGNs, we argue that they are not significant (within $1\sigma$ uncertainty). Also, adopting a stricter merger selection, we observe a $>1$ merger excess, indicating a high merger fraction in AGNs than non-AGNs. The numbers of galaxies and fractions for each group are listed in Table \ref{tb3}.

The overall merger fraction is higher in the GAMA sample than the SDSS sample, which could be due to the deeper imaging of KiDS revealing subtle features, higher redshift range in the GAMA sample, and/or difference in the training sample used in the CNN. 


\subsection{Merger fraction dependence on stellar mass, bolometric luminosity and obscuration}

In order to investigate the dependence of merger fraction on stellar mass, we separate AGNs into different mass bins for the SDSS sample, for which stellar masses are derived using the methods described in \citet{Kauffmann03} and \citet{Salim}. For the GAMA sample, in addition to stellar mass bins we also divide the sample into three redshift bins for MIR AGNs and two redshift bins for OPT AGNs (same as Section \ref{r1}). The stellar masses of the GAMA sample are derived from spectral energy distribution (SED) fitting \citep{Taylor11}. By selecting galaxies that appear in both the GAMA and SDSS samples we confirm that stellar masses derived through two different methods do not have a significant difference, with a median $\rm M_{*,SDSS}-M_{*,GAMA}$ of 0.07 dex.

\begin{figure*}
\includegraphics[width=\linewidth]{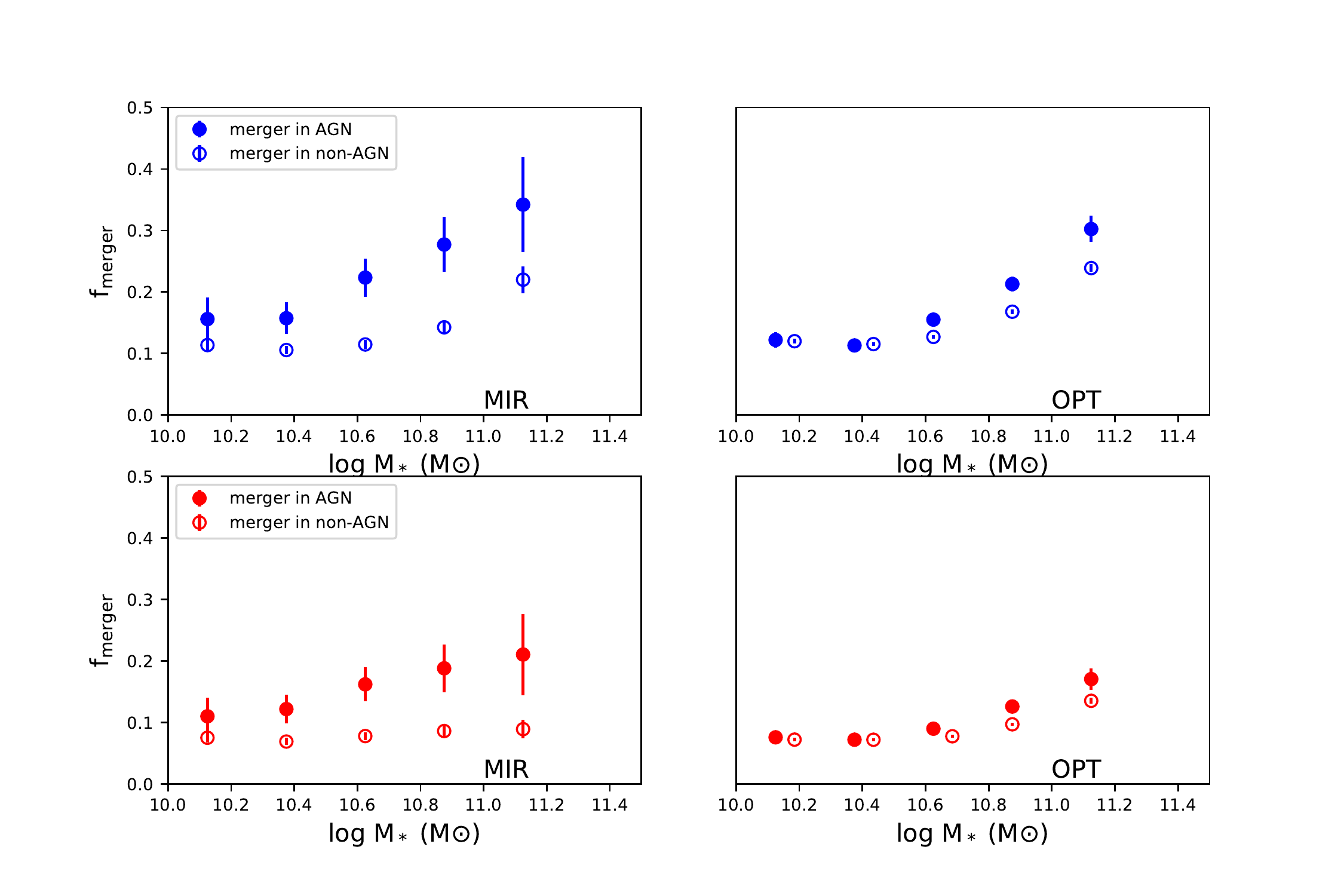}
\caption{Top: For the SDSS MIR AGNs (left) and OPT AGNs (right), the merger fraction increases as stellar mass increases in the mass complete regime. Bottom: Same as the top panel, but with stricter merger identification. Errors are calculated through binomial statistics. }
\label{merger_in_AGN_mass_sdss}
\end{figure*}

\begin{figure*}
\includegraphics[width=\linewidth]{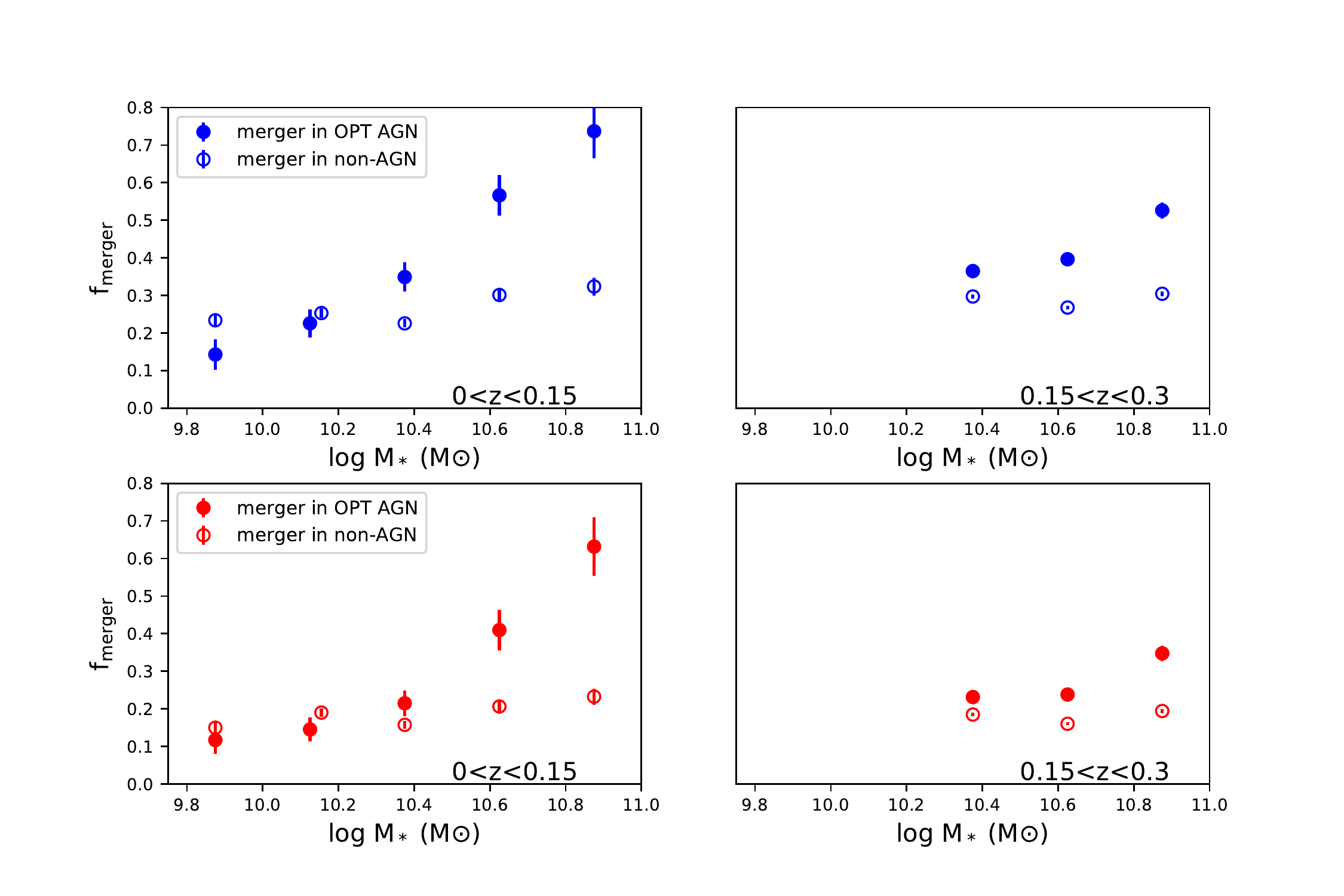}
\caption{ Top: For the GAMA OPT AGNs, the merger fraction increases as stellar mass increases in the mass complete regime in different redshift bins (indicated in the bottom right corner in each panel). Bottom: Same as the top panel, but with stricter merger identification. Errors are calculated through binomial statistics. We find an increase of the merger fractions in AGNs as stellar mass increases.}
\label{merger_in_AGN_mass_gama}
\end{figure*}

Figure \ref{merger_in_AGN_mass_sdss} shows the comparisons of main merger fractions and stricter merger fractions in MIR and OPT AGNs as a function of the stellar mass for the SDSS sample in the mass complete regime (above the lowest mass at the highest redshift), compared to the merger fractions in the non-AGN control samples. We can clearly observe an increase in the merger fraction in AGNs as stellar mass increases, suggesting that AGNs in more massive hosts are more likely to undergo a merger event \citep[e.g.,][]{Hopkins08, Ellison19}. The positive trend with increasing stellar mass is weaker or non-existent in the non-AGN controls. Figure \ref{merger_in_AGN_mass_gama} shows the comparisons of main merger fractions and stricter merger fractions in the OPT AGNs as a function of stellar mass as well as redshift for the GAMA sample in the mass complete regime respectively, in comparison with the non-AGN control samples. We do not show plots for the MIR AGNs because it lacks a large enough sample. For the GAMA sample, we can observe a similar trend for the merger fraction in the OPT AGNs, while for control samples we find a more flat trend as stellar mass increases. However, the GAMA sample and SDSS sample are not identical in terms of merger identification. Nonetheless, our analysis supports the idea that mergers are more important in triggering AGNs hosted by more massive galaxies.

\begin{figure*}
\begin{subfigure}{.5\textwidth}
\centering
\includegraphics[width=\linewidth]{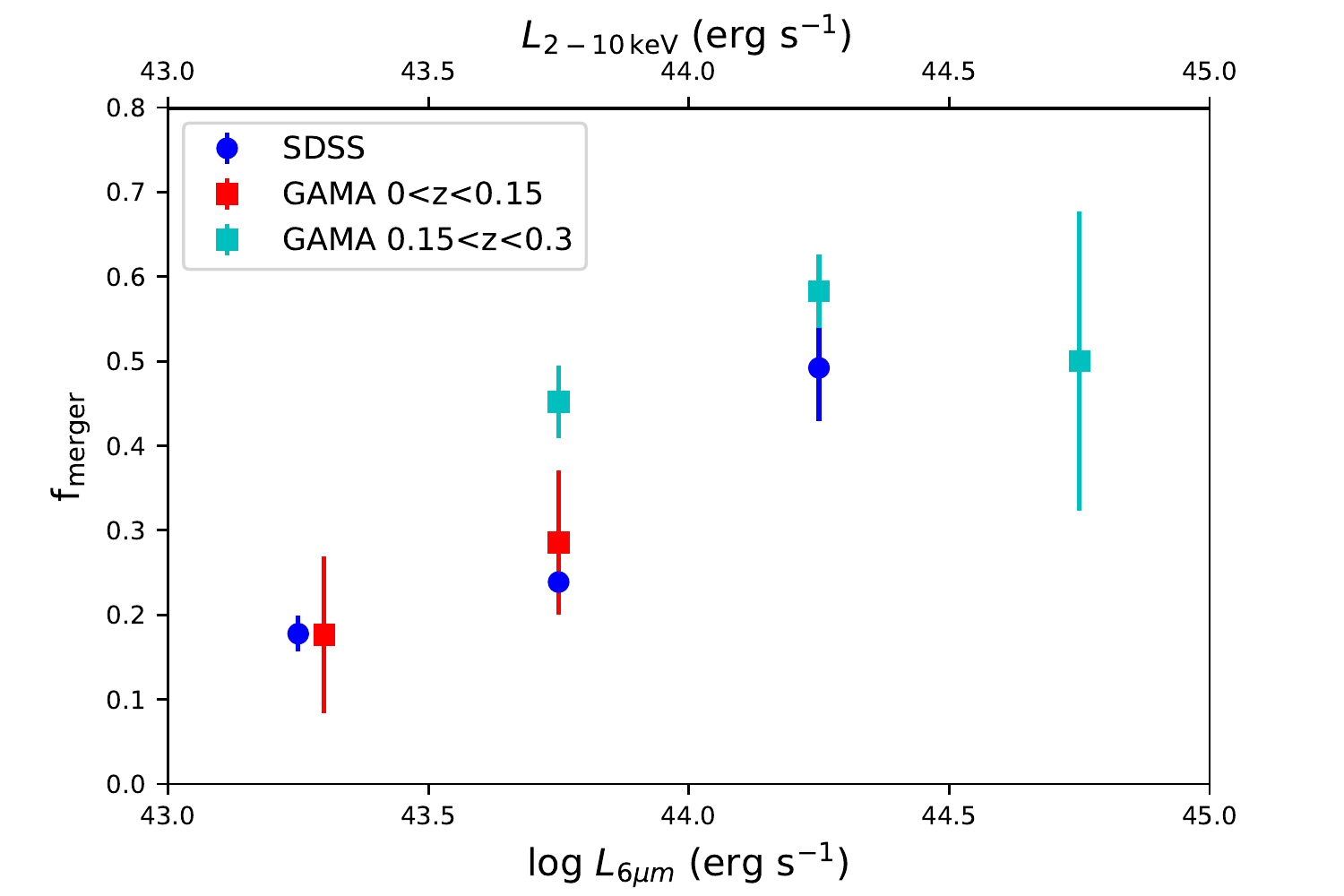}
\end{subfigure}
\begin{subfigure}{.5\textwidth}
\centering
\includegraphics[width=\linewidth]{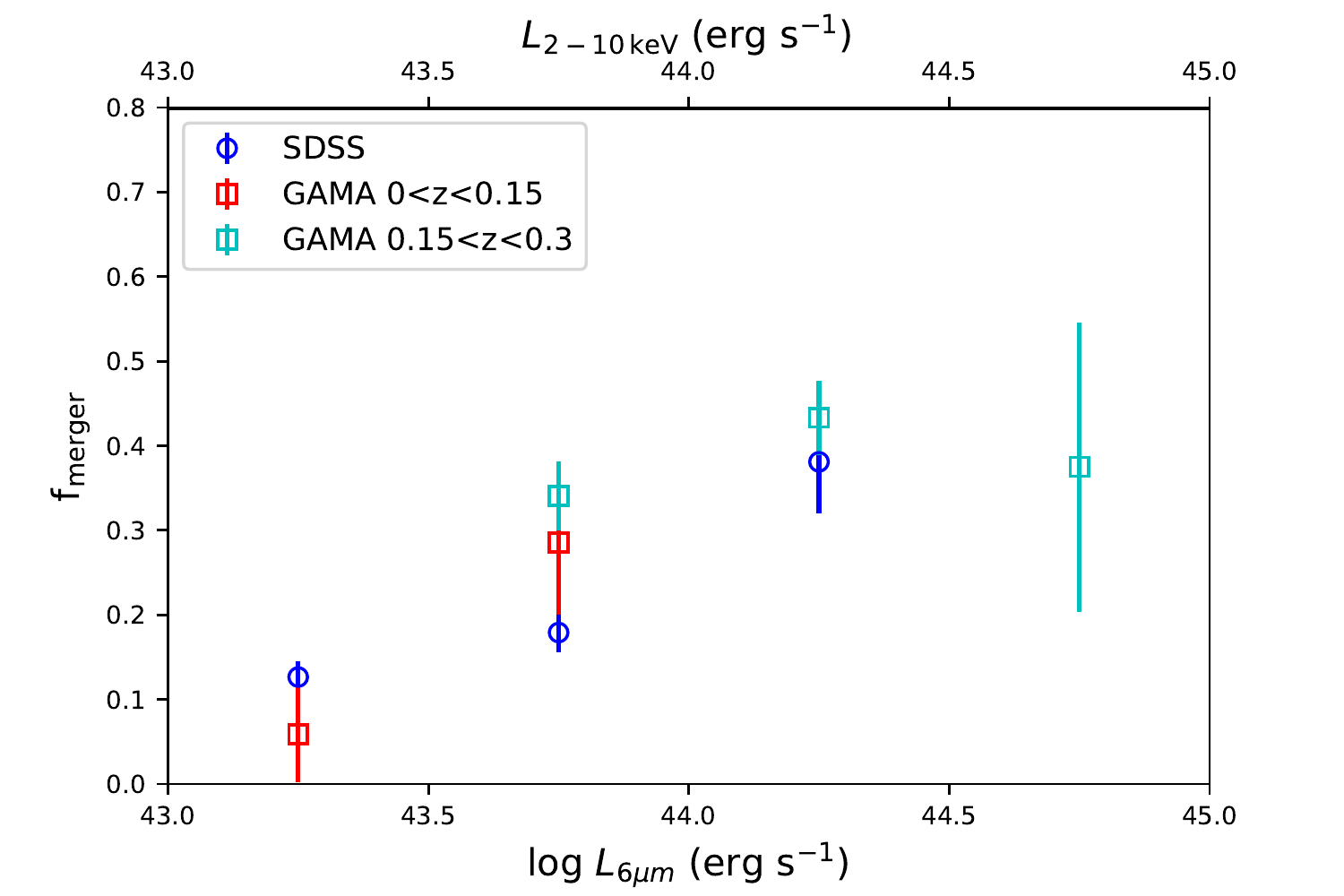}

\end{subfigure}
\caption{Left: The distributions of merger fractions in MIR AGNs as rest-frame $6\,\mu\text{m}$ luminosity increases for the SDSS sample and the GAMA sample. For the GAMA sample we also separate them into three redshift bins but the highest redshift bin is not shown due to small sample size. Right: same as left panel, but with stricter merger identification. Errors are calculated through binomial statistics.}
\label{L6}
\end{figure*}

\begin{figure*}
\begin{subfigure}{.5\textwidth}
\centering
\includegraphics[width=\linewidth]{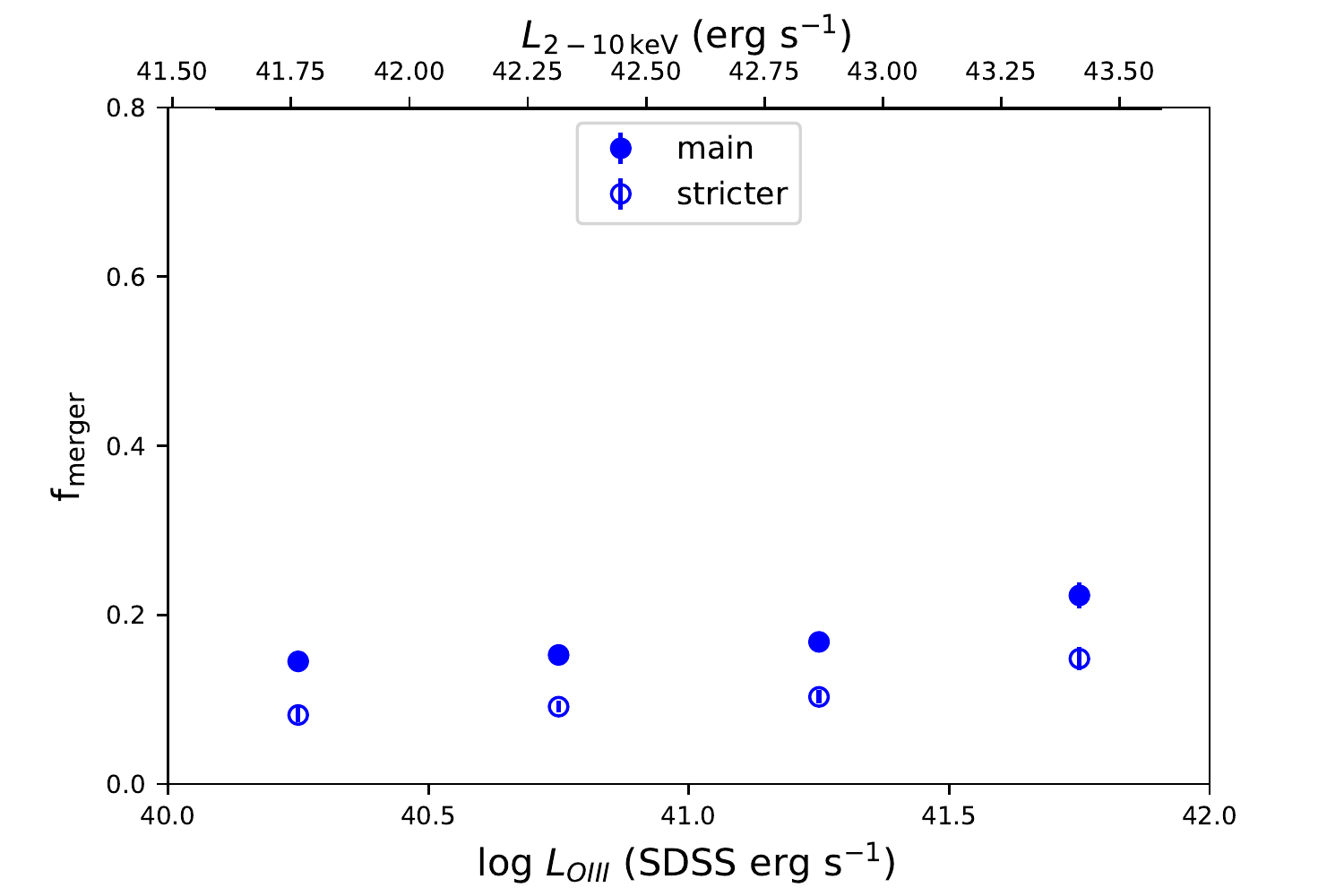}
\end{subfigure}
\begin{subfigure}{.5\textwidth}
\centering
\includegraphics[width=\linewidth]{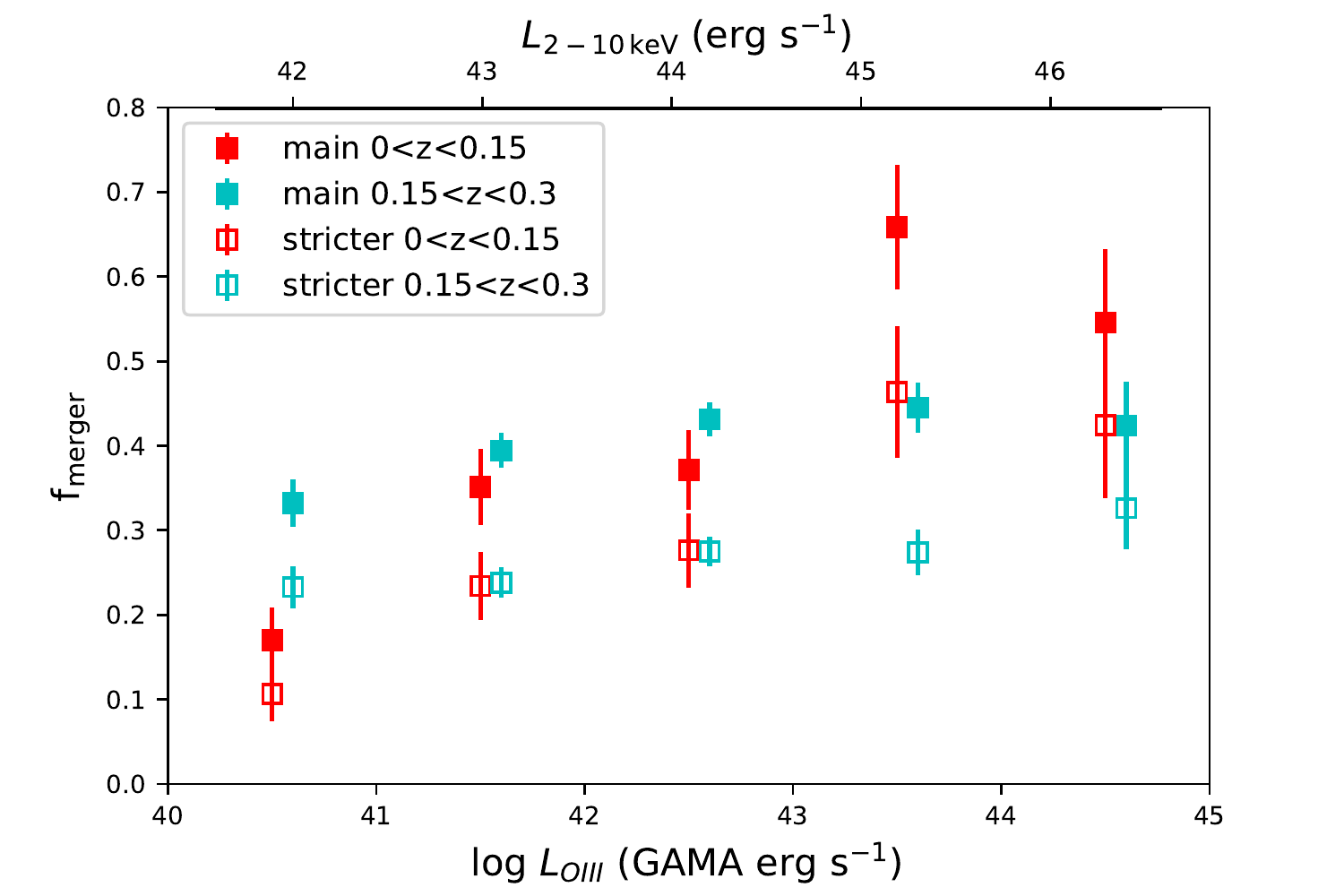}
\end{subfigure}
\caption{The distributions of main and stricter merger fractions in OPT AGNs as [OIII] luminosity increases for the SDSS sample (left) and the GAMA sample (right). For the GAMA sample we also separate them into two redshift bins. Errors are calculated through binomial statistics. The $x-$axes are indicated due to different aperture correction and flux calibration methods applied in the SDSS and GAMA surveys.}
\label{LO3}
\end{figure*}

We also split the AGNs into bins of bolometric power to investigate if there exists any dependence. Some studies found a higher occurrence of mergers in more luminous AGNs \citep[e.g.,][]{Hasinger08, Rosario, Santini12, Ellison19}, while others did not \citep[e.g.,][]{Hewlett, Villforth17}. We use the rest-frame $6\,\mu\text{m}$ luminosity and [O III] luminosity to represent bolometric AGN luminosity for MIR AGNs and OPT AGNs respectively. Similar to Section \ref{r1}, for the GAMA sample we also divide the sample into three redshift bins for MIR AGNs and two redshift bins for OPT AGNs. Figure \ref{L6} and \ref{LO3} show merger fractions as  AGN bolometric luminosity increases in the complete regime (lowest luminosity at the highest redshift). 
We do not include non-AGN control samples for a comparison because the rest-frame $6\,\mu\text{m}$ luminosity and [O III] luminosity are only relevant for AGNs. Due to difference in the aperture correction and flux calibration methods applied in the SDSS survey and the GAMA survey \citep[see][]{Brinchman, Hopkins13}, we denote [O III] luminosity as SDSS [O III] luminosity for the SDSS sample in the left panel of Figure \ref{LO3} and GAMA [O III] luminosity for the GAMA sample in the right panel of Figure \ref{LO3} respectively.

For MIR AGNs, the SDSS sample and the GAMA sample at $0<z<0.3$ show an increase in merger fraction as bolometric luminosity increases, growing by a factor of more than 2 from low to high bolometric luminosity. The GAMA sample at the highest redshift bin is not shown due to small sample size above the completeness limit. Due to the degeneracy between stellar mass and AGN bolometric luminosity, we cannot rule out the possibility that this increasing trend is led by the increasing trend between merger fraction and stellar mass. Due to limited statistics, we cannot distinguish whether stellar mass or AGN power is the intrinsic factor that drivers the merger fraction in AGNs.

For OPT AGNs, the situation is more complex. We observe a nearly flat trend for the SDSS sample. For the GAMA sample in the lower redshifts, we can see a clear gap of merger fraction between less powerful and more powerful AGNs, while for those in the higher redshifts, we observe a flat trend. Similarly, \citet{Ellison19} found a slight increase ($\sim 5\%$) in merger fraction at $40<L_{\rm [O III]}<42 \rm \, erg s^{-1}$ and an obvious enhancement at $L_{\rm [O III]}>42 \rm \, erg s^{-1}$.

\begin{figure}
\includegraphics[width=\linewidth]{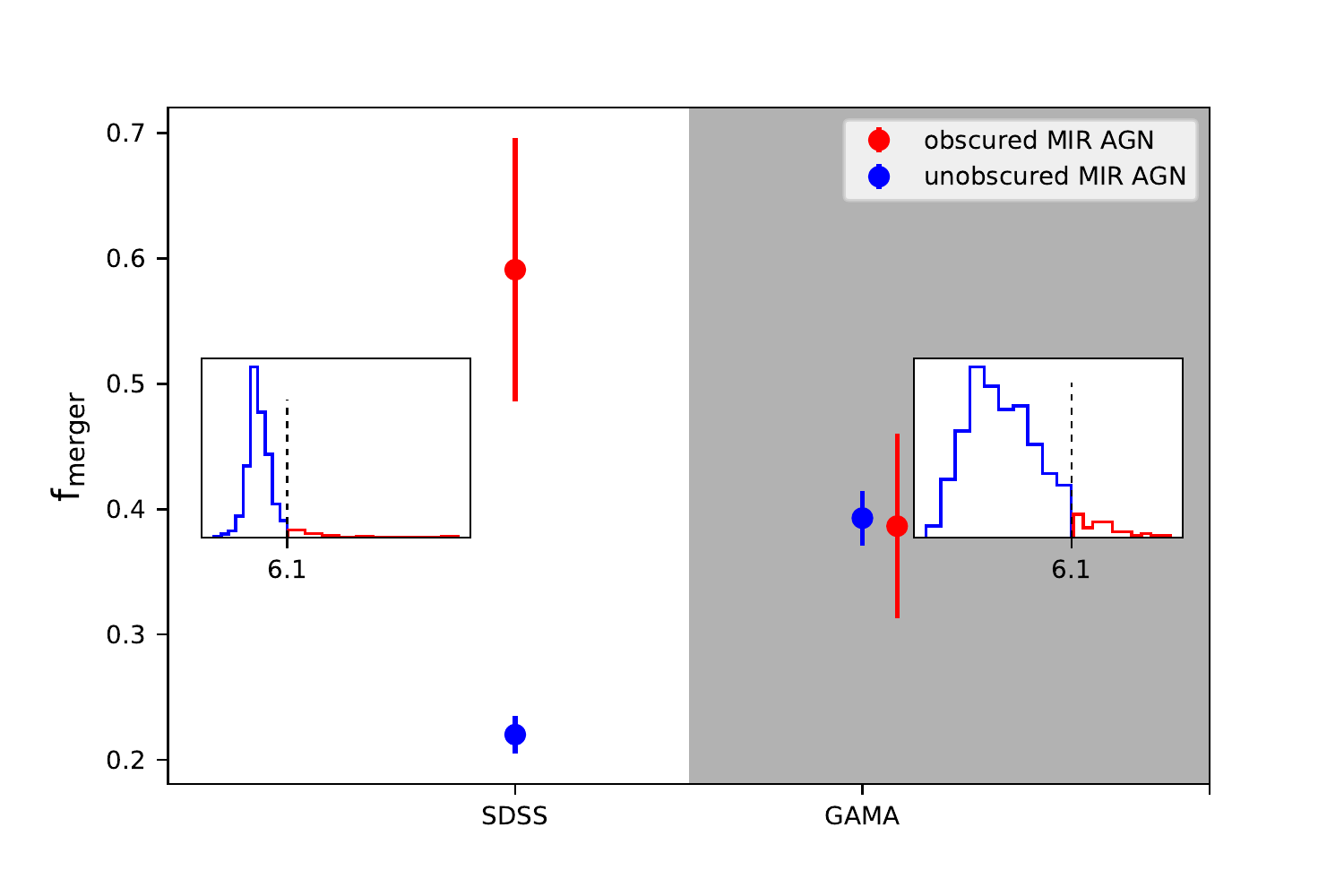}
\caption{Merger fraction in obscured/unobscured MIR AGNs for the SDSS and GAMA samples. The histograms of $r-W2$ color for each type are inserted. Errors are calculated through binomial statistics.}
\label{type12}
\end{figure}

For the MIR AGNs, we also take advantage of their optical-IR color to split them into unobscured and obscured AGNs. We adopt \citet{Hickox07} criterion $m_{\text{R}}-m_{4.5\,\mu\text{m}}=6.1$ (in Vega magnitude) using the SDSS $r$-band photometry and WISE $4.6\,\mu\text{m}$ photometry. Figure \ref{type12} shows the merger fractions in obscured/unobscured MIR AGNs for the SDSS sample and the GAMA sample. Obscured AGNs in the SDSS sample are more likely to be hosted in mergers than unobscured AGNs, despite the large uncertainty due to small number of obscured AGNs (22 SDSS MIR AGNs are obscured). $59.09\pm{10.48}\%$ of obscured AGNs in the SDSS sample are mergers while $22.01\pm{1.49}\%$ of unobscured AGNs in the SDSS sample are mergers. This higher fraction of mergers in obscured AGNs is consistent with previous studies in the IR and X-ray bands, e.g., early studies on LIRGs and ULIRGs \citep[e.g.,][]{Sanders96, Veilleux02}, hot dust-obscured galaxies \citep[e.g.,][]{Fan}, and heavily obscured X-ray AGNs \citep[e.g.,][]{Kocevski15}. The GAMA sample lacks significant contrast, possibly due to the fact that the majority ($\sim 85\%$) of the GAMA obscured AGNs are at $z>0.2$, which might raise difficulty in identifying mergers. When we limit to the lowest redshift bin, $50.00\pm{25.00}\%$ of obscured AGNs are mergers while only $23.08\pm{5.84}\%$ of unobscured AGNs are mergers, supporting that obscured AGNs are more likely to reside in mergers compared to unobscured AGNs.

\subsection{Merger fraction in low accretion rate AGNs}

\begin{figure*}
\includegraphics[width=\linewidth]{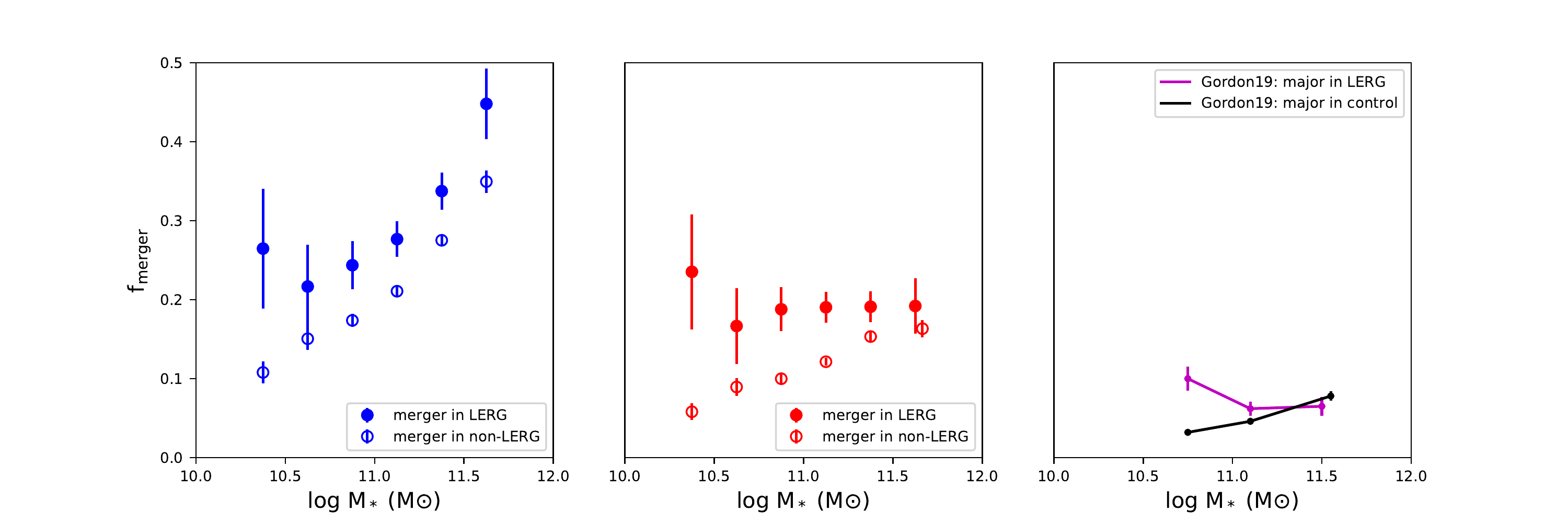}
\caption{Left: Merger fraction in LERGs/non-LERGs as a function of stellar mass. Middle: same as left panel, but with stricter merger identification. Errors are calculated through binomial statistics. Right: Minor and major merger fraction in LERGs and controls for the lowest quartile, interquartile and upper quartile of the LERG sample from \citet{Gordon19}.}
\label{lerg}
\end{figure*}

We find a merger fraction of $30.26\pm{1.31}\%$ in LERGs \citep[comparable to][]{Gordon19} and $22.94\pm{0.38}\%$ in non-LERGs which are higher than the merger fractions in the SDSS MIR and OPT AGNs and controls, suggesting that mergers still contribute to triggering these low accretion rate AGNs. In Figure \ref{lerg} we split LERGs and non-LERG controls into different stellar mass bins and find a positive trend as stellar mass increases, supporting an increasing importance of mergers in more massive AGNs. If we limit to stricter merger identification we observe a similar trend to that in \citet{Gordon19}, which found no difference of merger fraction in LERGs relative to non-LERGs in the most massive bin. \\

We summarize our findings as below:
\begin{description}
 
  \item[$\bullet$] We find a higher AGN fraction in mergers than in non-merger controls, suggesting that mergers do trigger AGNs.
  \item[$\bullet$] We find a higher merger fraction in AGNs than in non-AGN controls, implying that mergers play a significant role in AGN triggering.
  \item[$\bullet$] We find a dependence of merger fraction on stellar mass as mergers become more important for massive AGN hosts.
  \item[$\bullet$] As AGN bolometric luminosity increases, merger fractions in MIR AGNs and OPT AGNs show different trends. Both methods show a high merger fraction in more powerful AGNs.
  \item[$\bullet$] We find a higher merger fraction in obscured AGNs than in unobscured AGNs, consistent with previous studies.
  \item[$\bullet$] We find that mergers also play a significant role in triggering LERGs. 
  
\end{description}

\section{Discussion}\label{discussion}
\subsection{Comparison with previous works}
In Section \ref{r1} we find that our sample shows a smaller contrast of AGN excess in mergers relative to non-mergers compared with previous studies. In Section \ref{r2} we also find that our work shows a smaller contrast of merger fractions in AGNs and non-AGNs, with a merger excess of up to $\sim1.5$ in AGNs relative to non-AGN controls. \citet{Ellison19} found that AGNs are $\sim 2$ times more likely to be hosted in mergers compared to non-AGN controls, for a sample of 1124 optically selected AGNs and 254 MIR selected AGNs. In addition to the differences of our sample discussed in Section \ref{r1}, \citet{Goulding} proposed that AGN activity can occur sporadically during the entire stage of merger event. During the first and second passage, non-AGN phase can last longer than AGN activity. When the galaxies approach each other and begin to coalesce, AGN activity becomes more long-lived. Visual inspection bias merger selection towards more obvious mergers, which are more likely to be found in association with AGN activity than non-AGN activity, leading to a significant merger excess in AGNs relative to non-AGNs. If we limit to stricter mergers the merger excess can be $\sim 2$, which is more consistent with previous studies.

Many studies reported no excess of morphological disturbances in AGN hosts compared to a control sample, mostly using X-ray detected AGNs at high redshifts \citep[e.g.,][]{Grogin05, Gabor, Cisternas, Kocevski12}. High redshift samples can be biased towards luminous quasars \citep{Mechtley, Villforth17} that outshine their host galaxies, making it harder to identify mergers, especially post-mergers. In addition, highly obscured AGNs in which soft X-ray photons are obscured can be missed, showing an excess of merger fraction that is hidden in other studies \citep{Kocevski15}.

Simulations predict that galaxy mergers are able to transport gas inwards, leading to accretion around the central black hole \citep{Springel2}, and produce more luminous AGNs that cannot be explained by stochastic fueling \citep{Hopkins14}. However, \citet{Draper} found that mergers are not the only triggering mechanism for all AGNs and non-mergers processes are the dominant triggering mechanism by AGN population synthesis modeling. Also, \citet{Steinborn} found that less than 20\% of AGN hosts at $z=0-2$ have experienced a recent merger. Our work finds an increase of merger fraction in AGNs that reside in more massive galaxies and are most powerful. The percentage of AGNs that are mergers relative to all AGNs is $16-40\%$, suggesting a significant but maybe not dominant role of mergers in AGN triggering.

\subsection{Merger sequence}\label{d2}
Previous studies proposed an evolutionary track in the merger process: when two galaxies have a close encounter, gas is funneled towards the central region, increasing the local surface density and triggering starbursts.. This large gas reservoir also fuels the nuclear accretion activity when they lose most of their angular momentum and fall into the vicinity of central black holes due to gravitational torques \citep[e.g.,][]{DiMatteo, Springel1, Springel2}. As merging proceeds and galaxies coalesce, most AGNs are obscured by circum-nuclear dust, making them look extraordinarily red. These dust enshrouded AGNs serve as an explanation for the question of why most ULIRGs show merging features in early studies \citep[e.g.,][]{Sanders96, Murphy96, Veilleux02}. After final coalescence, when AGNs eventually expel the surrounding dust \citep[e.g., through AGN feedback,][]{Springel2}, they outshine the host galaxies and become optically visible, resulting in unobscured AGNs \citep{Sanders88, Hopkins06, Kocevski15}. 



In terms of the dependence of merger fraction on AGN bolometric luminosity, we observe an increasing trend in MIR selected AGNs, while for OPT AGNs we find a more flat trend at lower luminosities and an enhancement in the higher luminosity regime in the lower redshift range. In the higher redshift range the enhancement regime may not be covered by the dynamical range of our sample. We speculate that MIR selected AGNs exist more in late stage mergers with a thick dust envelope \citep[see illustration in][]{Kocevski15}. The more powerful AGN luminosity caused by more violent accretion bringing more gas supply could be easily recognized by a more disturbed morphology, leading to more merger identifications as luminosity increases. Or this increasing trend between merger fraction and AGN power is simply a by-product of the increasing trend between merger fraction and stellar mass. We do not have enough sample to figure out which factor is intrinsic. While OPT selected AGNs occur more in early stages or post-merger stages in which the dust is not compact enough to enshroud the nuclei or already expelled. The level of disturbance is not as clear as that in late stage mergers, leaving a relative flat trend. Those with higher luminosity could be in a transitioning merger phase from early stage to late stage, showing more disturbed morphology and having more matter supply supporting rapid accretion. 

Furthermore, the merger evolutionary scenario predicts that obscured AGNs are more likely to be hosted in mergers than their unobscured counterparts \citep{Satyapal, Kocevski15, Weston}, which can be seen from the comparison of merger fractions in obscured and unobscured MIR AGNs. 


\subsection{Caveats}\label{caveats}
Despite the high accuracy when identifying mergers using CNNs, the overall merger fraction in all galaxies is low, leading to contamination of non-mergers in the mergers. Assuming 1000 galaxies in which 10 $\%$ are real mergers, even a 90$\%$ accuracy will cause 100 galaxies to be incorrectly identified. In an extreme situation where these 100 galaxies are all non-mergers misidentified as mergers, then the final CNN-identified merger sample will include 200 galaxies (100 real mergers plus 100 misidentified galaxies), causing half of the sample contaminated by non-mergers. This contamination exists in both the AGN sample and the non-AGN control sample. 

\begin{figure}
\includegraphics[width=\linewidth]{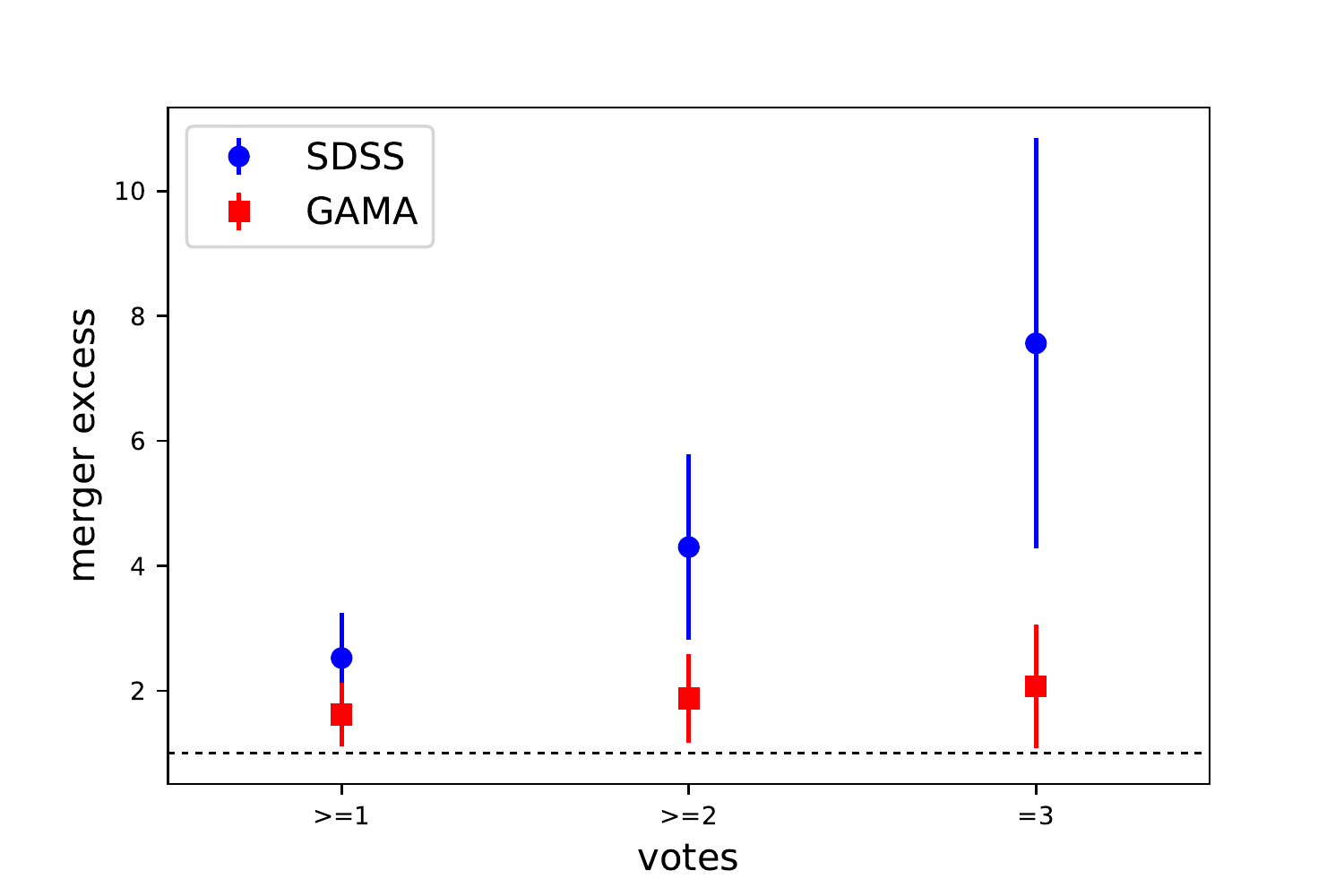}
\caption{Merger excess in MIR AGNs relative to non-AGN controls for the SDSS sample and the GAMA sample. The $x$-axis indicates the number of positive votes based on visual inspection.}
\label{visual}
\end{figure}

To assess the influence of this contamination and given the fact that it would be extremely time-consuming to visually inspect all the mergers, we only visually inspect the 184 and 213 mergers in the MIR AGNs of the SDSS sample and the GAMA sample respectively. For the controls we also visually inspect the 1200 mergers in the non-MIR AGNs of the SDSS sample and randomly select 600 mergers from the entire 1560 mergers in the non-MIR AGNs of the GAMA sample. The visual inspection is done independently by FG, LW and WJP. By selecting all mergers that have more than one, two and all three positive votes, in Figure \ref{visual} we find an increase of the merger excess in MIR AGNs relative to non-AGN controls for the SDSS sample and the GAMA sample.

There are advantages and disadvantages in terms of merger identification methods using CNN and visual inspection. Our CNN merger identification method is affected by non-merger contamination, but it can work efficiently on a large sample. Visual inspection, even though it is not always reliable, is less likely to be affected by contamination, but it would bias towards more obvious mergers, and is very time-consuming. Despite the non-merger contamination, our results are in qualitatively agreement compared to previous studies using visual inspecting in identifying mergers. In addition, when applying a stricter merger threshold and visually inspecting a smaller sample of mergers, our results are more consistent.


\section{Conclusion}
We select the SDSS DR7 spectroscopic data at redshifts $0.005<z<0.1$ and the GAMA spectroscopic data at redshifts $0<z<0.6$. We take advantage of deep leaning convolutional neural networks to identify mergers based on SDSS and KiDS images. We adopt two methods to classify AGNs, a WISE MIR color cut and optical emission line diagnostics, totaling 799 MIR AGNs and 5420 OPT AGNs for the SDSS sample, 543 MIR AGNs and 2750 OPT AGNs for the GAMA sample. We also select LERGs to analyze the connection between mergers and low accretion rate AGNs. We build a strictly matched control sample for each subgroup to investigate the connection between mergers and AGNs. Our findings are as follows:

(1) In terms of AGN fraction in mergers compared to non-mergers, AGNs are more likely to be found in mergers than in non-mergers, with a comparison of $1.63\pm{0.05}\%$ vs $1.45\pm{0.02}\%$ ($0.34\pm{0.02}\%$ vs $0.20\pm{0.01}\%$) for the SDSS OPT (MIR) AGNs and controls, $3.63\pm{0.11}\%$ vs $2.53\pm{0.03}\%$ ($0.73\pm{0.05}\%$ vs $0.53\pm{0.01}\%$) for the GAMA OPT (MIR) AGNs and controls, suggesting that mergers are able to trigger nuclear activity.
 
(2) $16.40\pm{0.5}\%$ ($23.03\pm{1.49\%}$) of the SDSS OPT (MIR) AGNs and $39.09\pm{0.93}\%$($39.23\pm{2.1}\%$) of the GAMA OPT (MIR) AGNs show merging features. The difference in the two samples may be attributed to the fainter detection limit of the KiDS imaging survey, high redshift range, and/or difference in the training samples of the CNN. Mergers play a significant role in triggering AGNs. Whether mergers dominate AGN triggering is still not confirmed considering the quality of the merger sample, the different timescales of merger events and AGN activity and so on.

(3)  At the same redshift, the merger fraction in AGNs increases as stellar mass increases, indicating that mergers are more important in triggering AGNs in more massive host galaxies.

(4) For LERGs which accrete at low rates we also observe a higher fraction of mergers than controls ($30.26\pm{1.31}\%$ vs. $22.94\pm{0.38}\%$).

(5) Merger fraction in MIR selected AGNs shows an increase as AGN power increases while we do not see a clear trend with AGN power for optically selected AGNs. In both selection methods, merger fraction is higher in more powerful AGNs. We interpret this phenomenon under a merger evolution scenario which is also supported by a higher merger fraction in obscured AGNs than non-obscured AGNs, selected according to their optical-WISE color. 


 \begin{acknowledgements} 
We thank Sara Ellison and David Alexander for suggestions and comments which helped to improve the paper.\\

Funding for the SDSS and SDSS-II has been provided by the Alfred P. Sloan Foundation, the Participating Institutions, the National Science Foundation, the U.S. Department of Energy, the National Aeronautics and Space Administration, the Japanese Monbukagakusho, the Max Planck Society, and the Higher Education Funding Council for England. The SDSS Web Site is http://www.sdss.org/.\\

The SDSS is managed by the Astrophysical Research Consortium for the Participating Institutions. The Participating Institutions are the American Museum of Natural History, Astrophysical Institute Potsdam, University of Basel, University of Cambridge, Case Western Reserve University, University of Chicago, Drexel University, Fermilab, the Institute for Advanced Study, the Japan Participation Group, Johns Hopkins University, the Joint Institute for Nuclear Astrophysics, the Kavli Institute for Particle Astrophysics and Cosmology, the Korean Scientist Group, the Chinese Academy of Sciences (LAMOST), Los Alamos National Laboratory, the Max-Planck-Institute for Astronomy (MPIA), the Max-Planck-Institute for Astrophysics (MPA), New Mexico State University, Ohio State University, University of Pittsburgh, University of Portsmouth, Princeton University, the United States Naval Observatory, and the University of Washington.\\

GAMA is a joint European-Australasian project based around a spectroscopic campaign using the Anglo-Australian Telescope. The GAMA input catalog is based on data taken from the Sloan Digital Sky Survey and the UKIRT Infrared Deep Sky Survey. Complementary imaging of the GAMA regions is being obtained by a number of independent survey programmes including GALEX MIS, VST KiDS, VISTA VIKING, WISE, Herschel-ATLAS, GMRT and ASKAP providing UV to radio coverage. GAMA is funded by the STFC (UK), the ARC (Australia), the AAO, and the participating institutions. The GAMA website is http://www.gama-survey.org/. Part of this work is based on observations made with ESO Telescopes at the La Silla Paranal Observatory under programme ID 177.A-3016.\\

This study is based on observations made with ESO Telescopes at the La Silla Paranal Observatory under programme IDs 177.A-3016, 177.A-3017, 177.A-3018 and 179.A-2004, and on data products produced by the KiDS consortium. The KiDS production team acknowledges support from: Deutsche Forschungsgemeinschaft, ERC, NOVA and NWO-M grants; Target; the University of Padova, and the University Federico II (Naples).

 \end{acknowledgements}

\bibliographystyle{aa}
\bibliography{draftbib}

\begin{appendix}
\section{Cutouts of mergers and non-mergers}
In this appendix we show cutouts of the same galaxies in the SDSS and KiDS imaging surveys.
Examples of merging galaxies are shown in the top four rows and non-mergers in the bottom four rows. As the KiDS survey is to $\sim 2.5$ magnitude deeper than SDSS imaging survey, subtle features are clearer in the KiDS images which can help merger identification.

\begin{figure*}
\includegraphics[height=18cm]{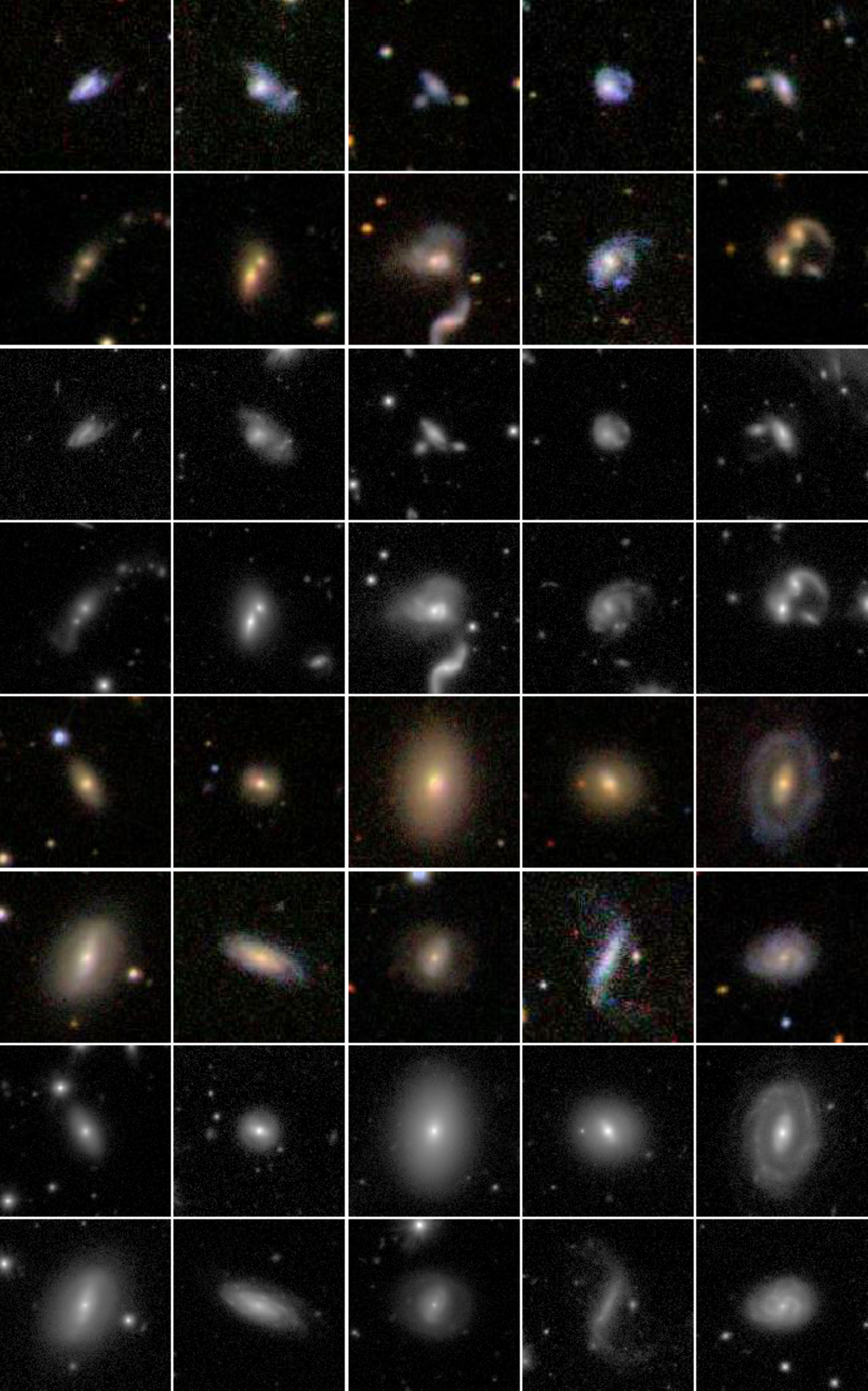}
\caption{Cutouts of the same mergers (top four rows) and non-mergers (bottom four rows). First two rows are SDSS mergers, second two rows are GAMA/KiDS mergers, third two rows are SDSS non-mergers and last two rows are GAMA/KiDS non-mergers. SDSS images are in $gri$ composite bands (first and third rows) with a size of 50.7$\arcsec$ $\times$ 50.7$\arcsec$. KiDS images are in $r-$band (second and fourth rows) with a size of 54.8$\arcsec$ $\times$ 54.8$\arcsec$. The depth of the KiDS imaging survey is $\sim 2.5$ magnitudes fainter than that of the SDSS imaging survey, revealing more subtle features.}
\label{cutouts}
\end{figure*}

\end{appendix}

\end{document}